\documentclass[11pt]{article}
\usepackage{amsfonts}
\usepackage{amsmath}
\usepackage{amssymb}
\usepackage{authblk}
\usepackage[superscript,biblabel,nomove]{cite}
\usepackage{gensymb}
\usepackage{graphicx}
\usepackage[hypertexnames=false]{hyperref}
\usepackage{xcolor}

\makeatletter
\renewcommand{\maketitle}{\bgroup\setlength{\parindent}{0pt}
\begin{flushleft}
  \textbf{\Large\@title}

  \@author
\end{flushleft}\egroup
}
\makeatother

\title{Identification of pressure points in modern power systems using transfer entropy}
\author[1,4,*]{Katerina Tang}
\author[2]{M. Vivienne Liu}
\author[1,3]{C. Lindsay Anderson}
\author[3]{Vivek Srikrishnan}
\affil[1]{Center for Applied Mathematics, Cornell University, Ithaca, NY, USA}
\affil[2]{Systems Engineering, Cornell University, Ithaca, NY, USA}
\affil[3]{Department of Biological and Environmental Engineering, Cornell University, Ithaca, NY, USA}
\affil[4]{Lead contact}
\affil[*]{Correspondence: kbt28@cornell.edu}
\date{\today}

\begin{document}

\maketitle

\section*{Summary}
Integration of variable energy resources---\emph{e.g.}, solar, wind, and hydro---and end-use electrification increase modern energy systems' weather-dependence.
Identifying critical infrastructure constraining the power grid's ability to meet electricity demand under weather-induced shocks and stressors is essential for understanding risks and guiding adaptation. 
We use transfer entropy to identify predictive pressure points: grid components whose utilization patterns provide early signals of downstream power shortages. 
We apply this method to simulations of New York State's proposed future grid under various meteorological and technological scenarios, showing that pressure points often arise from complex, system-wide interactions between generation, transmission, and demand. 
While transfer entropy does not support conclusions about causality, the identified pressure points align with known bottlenecks and offer insight into failure pathways. 
Furthermore, these pressure points are not easily predicted by high-level scenario features alone, underscoring the need for holistic and adaptive approaches to reliability planning in power systems with intermittent resources.

\section*{Introduction}
The reliability of the electric power system is being impacted by the increasing penetration of variable and intermittent sources of energy generation---such as solar, wind, and hydro---and rising electricity demand due to end-use electrification.
Greater penetration of variable generation and electrification increasingly couple the electric power system---across supply, transmission, and demand---with weather, leaving electricity reliability more susceptible to meteorological variability.\cite{craig_overcoming_2022,liu_heterogeneous_2023,grochowicz_using_2024,xu_resilience_2024} 

Given the interconnected nature of the electric power system, shocks from natural stressors across scales (including storms, heatwaves, and longer-scale hydroclimatic trends) can propagate to affect the entire grid.\cite{webster_integrated_2022,bernstein_power_2014,shuvro_impact_2018}
Due to the complexity of the power system, it is difficult to predict the dampening or amplification of shocks and where the greatest impacts will be felt.\cite{hines_cascading_2017,yang_small_2017} 
For example, local decreases in generation from natural resource variability or deratings due to temperature changes result in compensating increases elsewhere, potentially leading to strained supply\cite{grochowicz_using_2024} or transmission line congestion\cite{webster_integrated_2022} distant from the original disturbance.

The propagation and amplification of shocks due to the complex topology of the electric power grid results in the emergence of ``pressure points."\cite{webster_integrated_2022} 
By ``pressure point," we refer to infrastructure that consistently influences the occurrence or severity of power shortages by constraining the system's ability to meet electricity demand. 
Identifying pressure points in the power grid is key to identifying risks and evaluating potential interventions to improve reliability, such as infrastructure investments or demand response.
As these pressure points may be geographically or topologically distant from both the original shock(s) and where electricity demand is unmet, they can be hard to identify from patterns of shortages or transmission line congestion. 

Furthermore, natural meteorological variability does not necessarily cause component outages but can stress or weaken the power grid by simultaneously reducing weather-dependent energy generation output, limiting transmission, and/or increasing electricity demand over longer periods.\cite{grochowicz_using_2024,otero_characterizing_2022} 
This is in contrast with acute failures of grid components, which is a common setting for previous work on identifying critical infrastructure in power grids.
Some previous approaches\cite{wang_electrical_2010,bompard_analysis_2009,liu_recognition_2018} identify critical components by measuring how easily power can be generated and rerouted to meet demand after the complete removal of a node or link. 
These analyses adapt classical network centrality metrics to incorporate relevant operational constraints or electrical parameters.
It is unclear whether these approaches are applicable in cases where components are fully operational but at reduced capacity.

Another class of approaches analyzes the statistical relationships between the topological properties of nodes or links and the dynamics of simulated cascading failures triggered by their removal.\cite{yang_small_2017,beyza_geodesic_2021} 
In cases where the capacities of individual components---perhaps multiple, spatially correlated components---are reduced, the impacts on the power grid are not as easy to trace or quantify, especially if other components can compensate to help meet demand and largely avoid power outages.

In this study, we identify potential pressure points by inferring directed interactions from supply and transmission infrastructure (including generators, transmission interfaces, and batteries) to a specific load bus. 
The inference is based on hourly time series data that reflect resource utilization and unmet demand in simulated power flow studies. 
Framing this problem as inferring network connections fits within a broader literature using data-driven techniques to discover underlying grid interactions that may not be evident from direct or proximal electrical connections\cite{yang_vulnerability_2017,hines_cascading_2017,nakarmi_critical_2020}. 

Our metric of functional network connectivity is transfer entropy\cite{schreiber_measuring_2000} (see Methods for more details). 
Transfer entropy is an information-theoretic measure that can detect directed, nonlinear relationships between components in a system.
It quantifies the predictive information transfer from one process, $X$, to another, $Y$, by measuring how much knowledge of $X$'s past improves predictions of $Y$'s future, \emph{beyond} what is already predictable from $Y$'s own past.
When the transfer entropy from $X$ to $Y$ is positive, it suggests that $Y$ depends on $X$ in a predictive sense; although this does not establish causality, it indicates that the dynamics of $X$ provide unique information about $Y$'s dynamics.
The transfer entropy is zero when the processes are completely independent or when $Y$'s past already contains all the relevant information in $X$'s past, i.e., the two processes are perfectly coupled (Figures~\ref{fig:basic_te_example}a-c).
As evidence of its wide applicability, transfer entropy has been used to infer effective network connections from time series data in neuroscience\cite{wibral_transfer_2014,vicente_transfer_2011}, climatology\cite{runge_escaping_2012}, and econometrics\cite{dimpfl_using_2013}.

\begin{figure}[hbtp]
    \centering
    \includegraphics[width=0.9\linewidth]{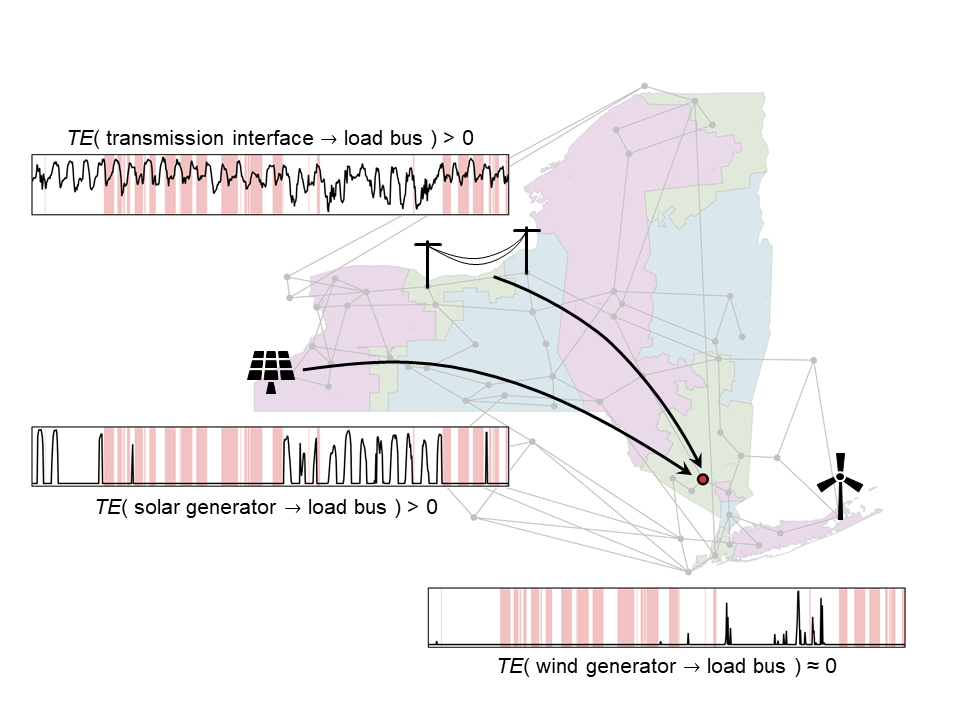}
    \caption{Identifying pressure points in a future New York State grid using transfer entropy. We focus on a single load bus (marked in red) and use transfer entropy to detect directed, predictive relationships between various components of the grid---wind, solar, and hydro generators; transmission interfaces; and batteries---and this load center. A component is a predictive pressure point if its pattern of utilization improves predictions of future power shortages at the target load bus. Here, interface flow and curtailment are plotted in black for a transmission interface and two generators, respectively, and hours with unmet demand at the target load bus are shaded in red.}
    \label{fig:workflow}
\end{figure}

We define a ``predictive pressure point'' for a given load bus as any infrastructure component---generator, transmission interface, or battery---for which there is statistically significant transfer entropy from the relevant utilization metric to the power shortage indicator at the load center of interest (Figure~\ref{fig:workflow}). 
This lets us identify---for example---a long-range predictive relationship between diurnal dynamics of transmission interface flow and downstream power shortages that a simple correlation analysis would miss.
And while curtailment patterns for solar and wind generators may appear similar, surrogate data testing can distinguish which generators provide statistically significant information about future load behavior.
We revisit similar examples in the results to demonstrate that a pairwise transfer entropy analysis can reveal subtle predictive interactions in the power grid that conventional analyses may overlook.

We demonstrate our approach using a proposed future New York State (NYS) power grid, as described and scoped by the state's Climate Leadership and Community Protection Act (CLCPA)\cite{new_york_state_climate_action_council_new_2022}. 
The CLCPA is an ambitious energy agenda that targets 70\% non-fossil electricity by 2030, 100\% zero-emission electricity by 2040, and eliminating emissions statewide by 2050. 
To meet these emission targets, the scoping plan calls for integrating additional variable and intermittent energy sources (\emph{e.g.}, on- and offshore wind, utility and behind-the-meter solar), upgrading and expanding transmission infrastructure, and electrifying end uses such as transportation and heating/cooling.\cite{new_york_state_climate_action_council_new_2022}
However, this plan exacerbates the spatial imbalance between energy supply, which is concentrated in upstate NYS (Figure~\ref{fig:map_zonal_caps}), and load, which is concentrated downstate, primarily in New York City.
Because of this imbalance---especially the uneven distribution of energy resources---NYS experiences complex responses to meteorological variability, making the future NYS grid a valuable case study for understanding the reliability risks of grids with intermittent resources.\cite{liu_heterogeneous_2023,kabir_quantifying_2024} 

This study employs simulations from ACORN (A Climate-informed Operational Reliability Network)\cite{liu_heterogeneous_2023}, a simplified optimal power flow model of the proposed 2040 NYS grid and connections to neighboring grids. These simulations were generated under scenario assumptions reflecting different meteorological forcings---primarily temperature increases---and  levels of variable energy penetration and electrification.
Hourly energy generation, transmission interface flow, battery state, and unmet demand are calculated for each meteorological-technological scenario by determining the optimal dispatch of resources to minimize power shortages (see Methods for details of the model, scenario design, and data-generating process).

In the first part of this study, we examine predictive pressure points in three specific scenarios representing distinct meteorological and technological challenges to the future NYS grid.
Previous studies of the reliability dynamics in future NYS grids\cite{kabir_quantifying_2024, liu_heterogeneous_2023} help us reason about the broad categories of pressure points that we might expect to see in each scenario.
These insights allow us to confirm that the specific predictive pressure points identified by transfer entropy largely align with known reliability challenges\cite{liu_heterogeneous_2023,kabir_quantifying_2024} and are consistent with expected system behavior.
Importantly, predictive pressure points can also reveal complex and sometimes unexpected interactions between generation and transmission patterns that ultimately lead to system failures, \emph{e.g.}, unmet demand. 
This highlights how interconnected the generation and transmission systems are in a grid with intermittent energy resources and how interventions aimed at maintaining reliability need to consider both simultaneously.

We then extend our analysis to all 6,600 meteorological-technological scenarios in the full ensemble, identifying predictive transmission pressure points in each and clustering the results to uncover shared system dynamics.
From a decision-making perspective, a key question is whether these pressure points can be anticipated from high-level indicators of grid conditions---\textit{e.g.}, temperature, resource availability, or installed capacity---so that planners can proactively design for reliability.
However, we find that these broad scenario features offer limited predictive power, reinforcing the conclusion of the preceding analysis: predictive transmission pressure points arise from complex, system-wide interactions among generation, transmission, and demand across spatial and temporal scales.
This unpredictability highlights the need for adaptive approaches and real-time system monitoring to support reliability planning in future power systems.

\section*{Results}
\subsection*{Heterogeneous grid dynamics across\\ meteorological-technological scenarios}
\label{subsec:narrative_scenarios}
\begin{figure}[hbtp]
    \centering
    \makebox[\textwidth][c]{
        \includegraphics[height=2.25in]{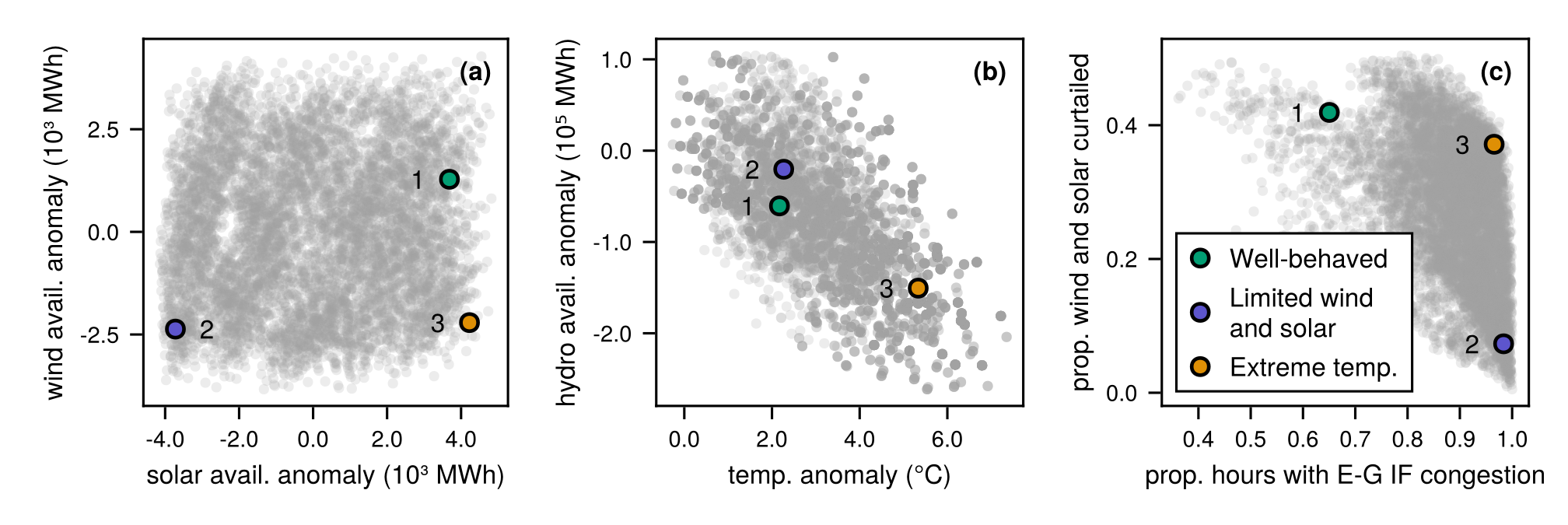}
    }
    \caption{Features of three chosen scenarios in the context of the full scenario ensemble. (a) Wind and solar availability and (b) hydro availability and temperature change are measured by the mean hourly---or, in the case of hydro availability, quarter-monthly---anomaly relative to a baseline scenario (see Methods). The three chosen scenarios also exhibit different levels of (c) wind and solar curtailment and upstate-to-downstate transmission congestion, reflecting heterogeneous reliability dynamics. We focus here on congestion in the E-G interface, a key interface in the upstate-to-downstate transmission pipeline.}
    \label{fig:scenario_features}
\end{figure}

One persistent challenge in using transfer entropy to identify networked impacts is that transfer entropy is a measure of network correlation, not causation.\cite{wibral_transfer_2014,bossomaier_introduction_2016} 
This means that predictive pressure points could hypothetically be the result of spurious correlations rather than meaningful causal patterns. 
To evaluate the plausibility of these inferences, we analyze three scenarios spanning a range of meteorological and technological assumptions (see Figure~\ref{fig:scenario_features}a-b) in more depth. 
In each case, we evaluate the identified predictive pressure points in the context of previous studies\cite{kabir_quantifying_2024,liu_heterogeneous_2023}, known structural features of the NYS grid, and the relatively simple power flow dynamics observed in ACORN.

A key structural feature of the CLCPA's proposed NYS grid is the spatial imbalance between energy supply and electricity demand, which ACORN simulations consistently demonstrate is a major challenge for grid reliability.\cite{kabir_quantifying_2024,liu_heterogeneous_2023}
To meet the zero-emissions generation target by 2040, the CLCPA scoping plan\cite{new_york_state_climate_action_council_new_2022} prescribes wind, solar, and hydro resource capacities for each of the eleven New York Independent System Operator (NYISO) load zones, labeled A through K (Figure~\ref{fig:map_zonal_caps}).
Most planned land-based wind, utility solar, and hydropower resources are located upstate---in zones A through E---and in zone F.
Meanwhile, electricity demand is disproportionately concentrated downstate in zones J and K, which have primarily offshore wind generation resources.
As a result, power typically flows from upstate to downstate, and the upstate-to-downstate transmission corridor---particularly the E-G interface---is a persistent bottleneck, including in the three meteorological-technological scenarios we consider here (Figure~\ref{fig:scenario_features}c).

We first examine a relatively ``well-behaved" scenario in which energy availability is high---largely because total installed wind and solar capacities are 126\% and 135\%, respectively, of CLCPA targets---and the assumed temperature increase is moderate (Figure~\ref{fig:scenario_features}a-b).
These favorable conditions reduce overall strain on the system, yet we see some pressure points still emerge.
Downstate power shortages occur exclusively during hours in which the critical E-G interface is congested.
However, this congestion does not appear to be a major barrier to the effective use of available energy resources: wind and solar curtailment upstate, congestion of the E-G interface, and unmet demand downstate co-occur in less than 9\% of simulated hours.
Instead, the large volume of curtailed wind and solar generation ($2.04\times 10^7$ MWh) compared to a much smaller total power shortage ($1.29\times 10^6$ MWh) points to a temporal mismatch between when these intermittent energy resources are available and when they are needed.

Next, we consider two more extreme meteorological-technological scenarios in which statewide energy availability is more constrained.
In the first of these two scenarios, wind and solar capacities are only 73\% and 63\%, respectively, of what is called for in the CLCPA scoping plan, but the temperature increase is still fairly moderate (Figure~\ref{fig:scenario_features}a-b); we refer to this scenario as the ``limited wind and solar" scenario.
The third and final scenario has relatively limited wind availability and a large temperature increase (Figure~\ref{fig:scenario_features}a-b). 
This extreme temperature increase depresses hydropower availability (Figure~\ref{fig:scenario_features}b, Figure~\ref{fig:load_and_hydro_vs_temp}a) and reduces transmission ampacity\cite{bartos_impacts_2016} while dramatically increasing overall demand due to increased cooling load (Figure~\ref{fig:load_and_hydro_vs_temp}b). 
Total statewide load over the entire simulation period is 121\% of the baseline (see Methods for details of the baseline scenario) with hourly load peaking at 147\% of the baseline.

The extreme conditions in the second and third scenario---particularly the limited offshore wind availability---result in a greater reliance on upstate energy supply to meet downstate demand.
This likely exacerbates congestion in the upstate-to-downstate transmission pipeline, forcing additional wind and solar curtailment and ultimately leading to power shortages\cite{liu_heterogeneous_2023} and increased system costs\cite{kabir_quantifying_2024}.
We also expect to see generation-side pressure points become more pronounced as constrained energy supply fails to keep up with electricity demand, particularly in the third, extreme temperature scenario.

\subsection*{Pressure points emerge from interactions between generation and transmission}
\label{subsec:ppps}

\begin{figure}[hbtp]
    \centering
    \makebox[\textwidth][c]{
        \includegraphics[height=2.0in]{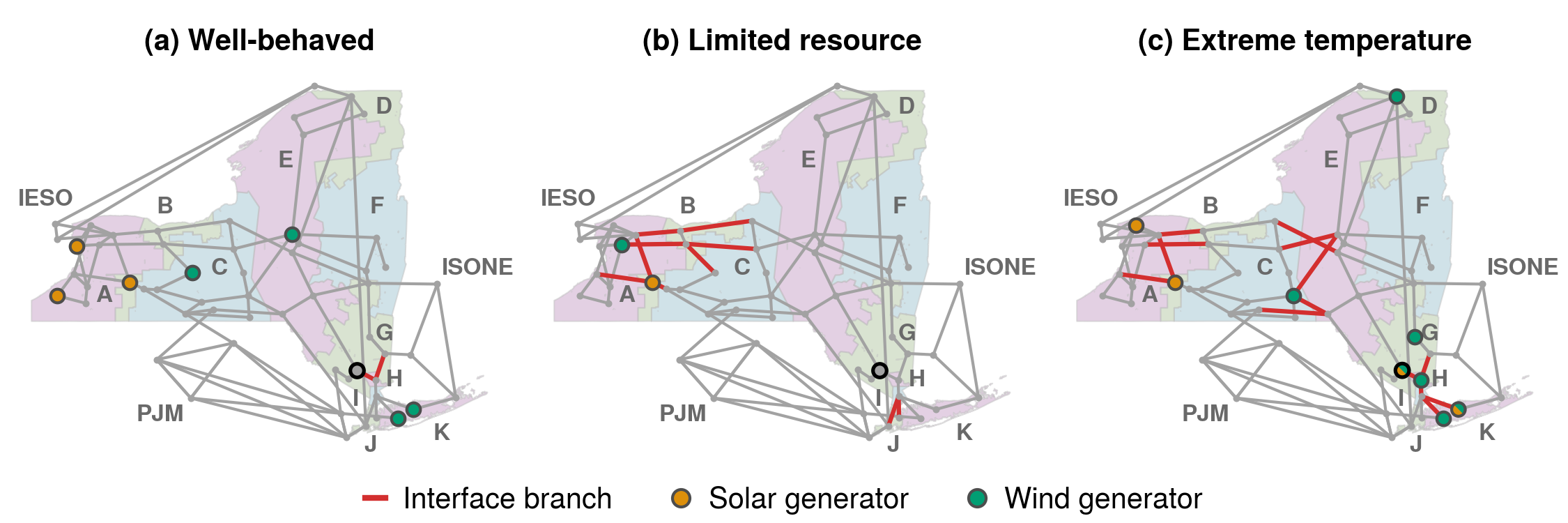}
    }
    \caption{Predictive pressure points identified by pairwise transfer entropy analysis for zone G load center (circled in black) across three chosen scenarios.}
    \label{fig:map_pressure_points}
\end{figure}

Throughout our transfer entropy analysis, we focus on identifying pressure points that help predict power shortages at a bus located in Zone G (highlighted in Figure~\ref{fig:map_pressure_points}). 
This bus lies along the critical (and frequently congested) upstate-to-downstate transmission pipeline and consistently exhibits the highest frequency and largest share of unmet demand across the entire scenario ensemble (Figure~\ref{fig:ls_hrs_and_prop_by_bus}). 
While electricity demand peaks further downstate, particularly in New York City and Long Island (zones J and K, respectively), shortages in those zones tend to co-occur with shortages at the zone G bus: in over 99\% of scenarios, at least 90\% of hours with unmet demand in zones J and K coincide with hours of unmet demand at this zone G location. 
This makes it a strategically informative site for identifying system-wide pressure points.

\begin{figure}[hbtp]
    \centering
    \makebox[\textwidth][c]{
        \includegraphics[height=2.8in]{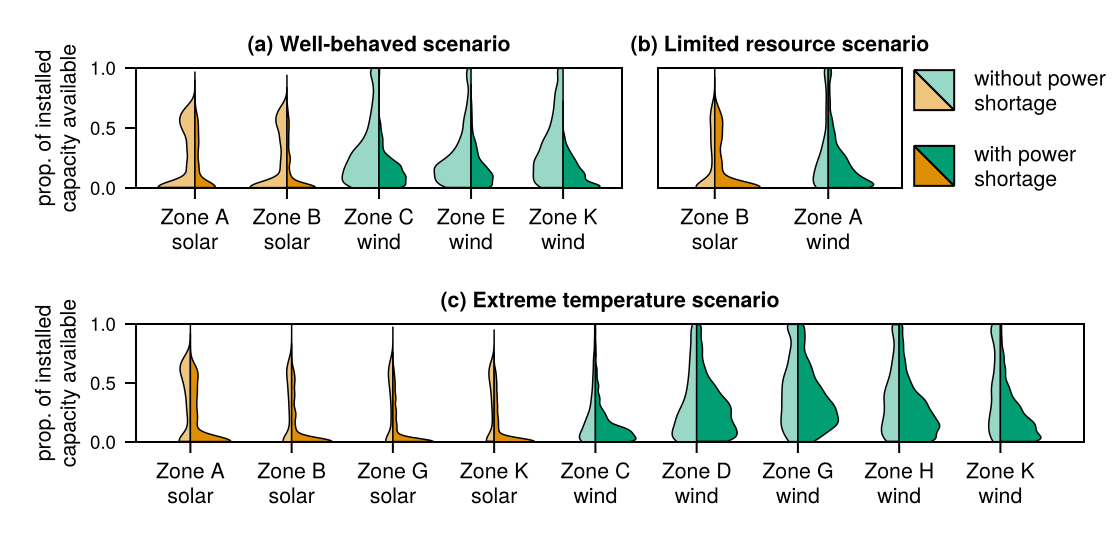}
    }
    \caption{Distribution of generation availability relative to installed capacity at identified predictive pressure points during hours with (right) and without (left) zone G power shortages across three chosen scenarios. All generation pressure points identified within a single zone are aggregated.}
    \label{fig:renew_ratios}
\end{figure}

In the well-behaved scenario, our transfer entropy analysis highlights temporal mismatches between intermittent energy generation and load as the primary drivers of power shortages.
Most predictive pressure points are generators where available wind and solar resources are consistently fully utilized---\emph{i.e.}, no curtailment occurs---during periods of unmet demand at the target load bus (Figure~\ref{fig:curtailment}a). 
Shortages are heavily concentrated during hours in which wind and solar availability at these sites drops below 50\% of nameplate generation capacity (Figure~\ref{fig:renew_ratios}a), particularly at night when solar is unavailable and when wind availability is also reduced.
These results reinforce the interpretation that timing, rather than supply capacity, is the key driver of supply-side pressure points in this scenario.

The G-H interface is also flagged as a predictive pressure point due to diurnal decreases in interface flow that are predictive of power shortages at the target load bus in zone G (Figure~\ref{fig:s140_GHif_zoneFrenewables}a). 
However, this predictive relationship is likely spurious: an underlying driver of unmet demand in zone G and G-H interface flow dynamics appears to be low solar availability in zone F (Figure~\ref{fig:s140_GHif_zoneFrenewables}b), which has the state's largest share of utility solar resources and contributes heavily to downstate energy supply.
Because the E-G interface is frequently congested (Figure~\ref{fig:scenario_features}c), solar generation in zone F and flow across the F-G and G-H interfaces become critical for meeting demand in zones G through K, amplifying the predictive role of the G-H interface.
We return to this issue of common drivers and the interpretability of transfer entropy in the Discussion.

Together, these findings underscore the need for expanded energy storage to address the temporal mismatch between energy supply and electricity demand. 
In over half of simulated hours, curtailment of wind and solar generation co-occurs with fully-met demand. 
Long-duration batteries as well as more flexible, longer-range options such as green hydrogen---which can time-shift surplus energy without exacerbating transmission line congestion\cite{liu_heterogeneous_2023}---could make more efficient use of the abundant but intermittent energy resources in this scenario.
These results are consistent with previous studies of grids with non-dispatchable, intermittent energy resources, which similarly emphasize the critical role of storage in achieving reliable grid operation.\cite{liu_heterogeneous_2023,kabir_quantifying_2024,grochowicz_using_2024}

Notably, the E-G interface is not identified as a predictive pressure point in this scenario despite being a known structural bottleneck. 
This is a consequence of its near-deterministic relationship with downstate power shortages in this scenario: unmet demand downstate is so strongly correlated with congestion of the E-G interface that the history of interface flow contains little unique predictive information.
We consider this specific limitation of transfer entropy as a metric of functional network connectivity in more depth in the Discussion.

In the second scenario we examine---the limited wind and solar resource scenario---our transfer entropy analysis also identifies supply-side predictive pressure points (Figure~\ref{fig:map_pressure_points}b).
However, here the predictive relationship reflects a more fundamental supply constraint: at flagged generators, wind and solar resources are again fully utilized during hours of unmet demand (Figure~\ref{fig:curtailment}b), but unlike the well-behaved scenario, power shortages are not limited to periods of extremely low energy availability (Figure~\ref{fig:renew_ratios}b).
This aligns with our expectations for this meteorological-technological scenario: underbuilt wind and solar generation capacity reduces system resilience.
More specifically, the lack of redundancy in generation capacity leaves the grid susceptible to ordinary meteorological variability such that even moderate decreases in wind or solar availability can lead to shortages.

Another consequence of underbuilt offshore wind in particular is the increased reliance on far-upstate energy resources, particularly the considerable hydropower and utility solar resources located in zone A.
As discussed previously, this exacerbates congestion along the upstate-to-downstate transmission corridor, and we see that the A-B and B-C interfaces are identified as predictive pressure points (Figure~\ref{fig:map_pressure_points}b).
These interfaces see extended periods of high utilization and congestion during shortage hours, averaging 90\% and 97\% of transmission capacity, respectively, compared to 77\% and 90\% during periods with no power shortage at the target bus. 
(See also Figure~\ref{fig:s69_ABif_BCif_IJif_utilization}a-b.)
Importantly, these interfaces are examples of geographically- and electrically-distant predictive pressure points arising from interactions between generation and transmission.
Transfer entropy's ability to surface these complex relationships---rather than just proximate or isolated drivers---is a key strength of our approach.

Taken together, these findings suggest that expanding either generation or transmission capacity alone is likely insufficient to address the underlying cause of reliability failures.
Without increasing the A-B and B-C interface transmission capacities, additional generation capacity in zones A and B could not be effectively utilized. 
In particular, the B-C interface is congested during 70\% of hours with unmet demand in zone G.
At the same time, increasing interface transmission capacity alone will not resolve shortages if there is insufficient generation capacity upstream.
Curtailment of wind and solar generation in zones A and B is rare, occurring in just 13\% of all simulated hours and only 4\% of hours with unmet demand in zone G, suggesting a fundamental supply shortfall.
Thus, capacity expansions must address both supply and delivery constraints.

The I-J interface is also identified as a predictive pressure point in this limited wind and solar resource scenario, reflecting a more subtle but still plausible causal connection.
Here, a consistent pattern of peaking interface flow from zone I to zone J---which has the largest electrical load among all NYISO zones---is predictive of unmet demand at the target load bus in zone G (Figure~\ref{fig:s69_ABif_BCif_IJif_utilization}c).
These peaks in interface flow largely occur during hours with low offshore wind availability in zone J (Figure~\ref{fig:s69_zoneJwind_IJif_scatter}a).
When offshore wind availability is high, shortages in zone J can largely be avoided---98\% of hours with unmet demand in zone J coincide with wind availability below 50\% of zone J's installed capacity---and the reverse flow of power from zone J to zone I can help meet demand elsewhere, including in zone K and indirectly in zone G (Figure~\ref{fig:s69_zoneJwind_IJif_scatter}b).
While power does not ever flow in the I~$\to$~H direction, \emph{i.e.}, from downstate to upstate, this pattern underscores the broader system-wide importance of downstate energy supply and transmission flexibility.
Notably, this is a relationship that a simple correlation analysis would likely miss; transfer entropy captures the predictive value of more complicated flow patterns---such as peaks---rather than steady levels or direction alone.

In the final scenario, extreme temperatures significantly increase electricity demand while decreasing hydropower availability; wind availability is also decreased (Figure~\ref{fig:scenario_features}a-b).
The grid becomes heavily reliant on solar generation---particularly from upstate zones and zone F---to meet this elevated load.
As a result, the predictive transmission pressure points observed in the previous scenarios are exacerbated (Figure~\ref{fig:map_pressure_points}c). 
For example, A-B interface congestion once again emerges as a strong predictor of downstate power shortages (Figure~\ref{fig:s290_ABif_CEif_GHif_HIif_IKif_utilization}a), occurring in 36\% of hours with unmet demand in zone G but only 7\% of hours with no unmet demand.
Diurnal fluctuations in interface flow are more pronounced here than in the two previously discussed scenarios, reflecting the system's increased dependence on zone A's solar generation in the absence of dispatchable hydropower.

More predictive transmission pressure points emerge along the upstate-to-downstate transmission pipeline---specifically at the C-E, G-H, and H-I interfaces (Figures~\ref{fig:s290_ABif_CEif_GHif_HIif_IKif_utilization}b-d).
As in the well-behaved scenario, these pressure points reflect a common driver: diurnal solar variability.
Lower interface utilization during periods of solar unavailability correspond to unmet demand in zone G.
When solar supply---particularly in zones C and F---are sufficient, these interfaces are more fully utilized and shortages are reduced.
This pattern illustrates that congested transmission can be primarily a consequence of power flow dynamics and does not necessarily indicate a bottleneck or limitation in the system but rather that available resources are being utilized appropriately to meet load.

Supply-side predictive pressure points are also widespread in this scenario, reflecting the broader challenge of meeting elevated demand. 
Many of these predictive pressure points are located at wind generators, and notably, power shortages occur even when wind availability reaches installed capacity (Figure~\ref{fig:renew_ratios}c).
This suggests that wind capacity---across the state---is underbuilt relative to the total load. 
Expanding wind generation is an obvious remedy, but where additional capacity is installed matters: adding more land-based wind upstate without upgrading transmission infrastructure risks exacerbating existing bottlenecks, while increasing offshore wind capacity downstate could directly alleviate local demand without relying on congested interfaces.

The need for these spatial considerations is further underscored by curtailment patterns.
At a solar generator in zone A flagged as a predictive pressure point, solar output is curtailed in about 8\% of hours with unmet demand in zone G (bottom panel of Figure~\ref{fig:curtailment}c). 
More broadly, we see curtailment somewhere in the system in about 64\% of hours with unmet demand, indicating that transmission constraints limit the effective use of available upstate energy resources.
These findings highlight the importance of coupling generation expansion with transmission upgrades.

The final predictive pressure point in this scenario is the I-K interface.
Here, periods of sustained flow in the in the K~$\to$~I direction predict power shortages in zone G (Figure~\ref{fig:s290_HIif_IKif_utilization}a).
This pattern reflects increased reliance on offshore wind from zone K to support demand in zone J, particularly when overall energy supply from upstate is reduced (Figure~\ref{fig:s290_HIif_IKif_utilization}b-c).
However, unlike the dynamics of the I-J interface observed in the previous scenario---where reverse flow helped mitigate zone G power shortages---here K~$\to$~I flow is insufficient to prevent zone G shortages. 
This is likely due to the relatively low interface flow limit in the K~$\to$~I direction (515 MW), which is much smaller than zone J's mean hourly load (approximately 8950 MWh).
Still, the dynamics of this pressure point are consistent with the broader system condition: underbuilt wind, strained solar supply, and overwhelming downstate demand.

\subsection*{Transmission pressure points are hard to predict based on meteorological and technological features}

In the previous subsection, we demonstrated that predictive pressure points reflect diverse and often complex system dynamics that can vary across meteorological and technological conditions.
This raises two decision-relevant questions.
First, which predictive pressure points---if any---are common across a broad range of meteorological-technological scenarios?
Second, can the emergence of specific pressure points be anticipated from broader operating conditions, allowing planners to anticipate and possibly address reliability risks in advance?
This process---identifying the underlying scenario features that result in similar system dynamics---is known as scenario discovery\cite{kwakkel_developing_2015} and can help guide decision-makers in planning for system reliability under uncertainty.

\begin{figure}[hbtp]
    \centering
    \makebox[\textwidth][c]{
        \includegraphics[height=1.75in]{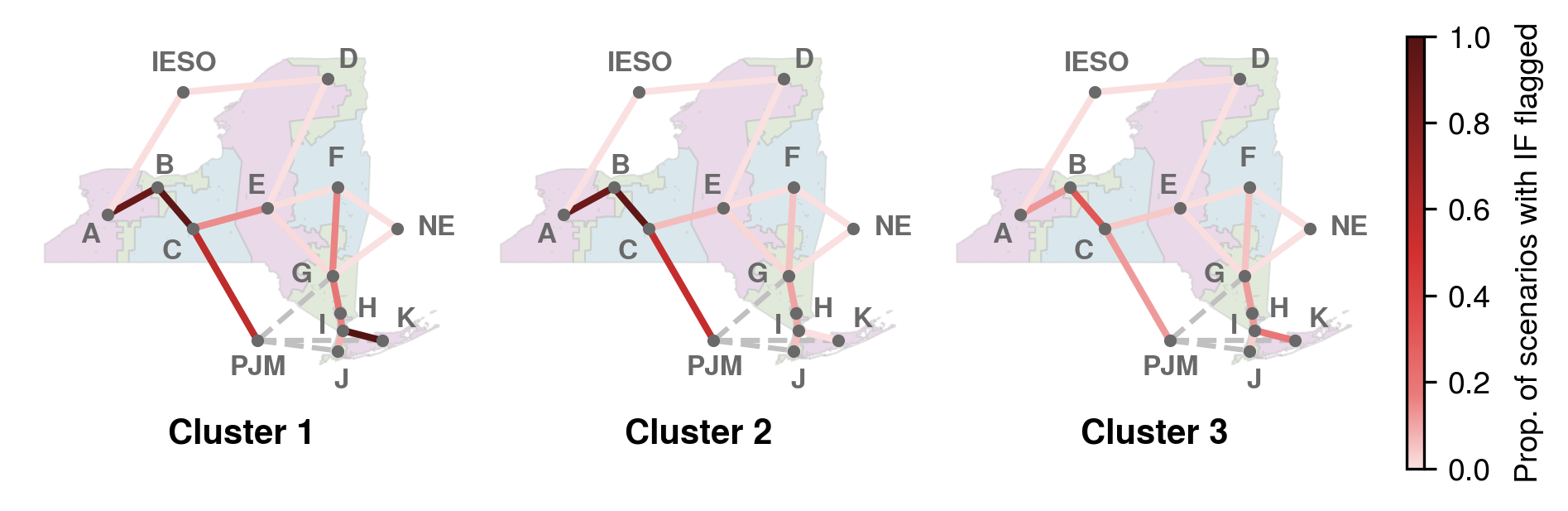}
    }
    \caption{Patterns of predictive transmission pressure points grouped into three clusters. Note that the NY-PJM interface connects zone C and the PJM ISO but does not include connections from the PJM ISO to zones G, J, and K, shown as dashed gray lines.}
    \label{fig:map_3_clusters}
\end{figure}

To investigate the first question, we group the 6,600 meteorological-technological scenarios based on similarities in their transmission-level predictive pressure points, which, as demonstrated above, capture critical reliability dynamics often overlapping with those suggested by generation-level predictive pressure points. 
A clustering analysis (see Methods for details) reveals three distinct clusters of scenarios, each characterized by a common spatial pattern of predictive transmission pressure points (Figure~\ref{fig:map_3_clusters}). 
That is, each cluster contains scenarios where similar subsets of transmission interfaces emerge as predictive pressure points, pointing to similar modes of system failure.
Scenarios in the first two clusters share multiple predictive pressure points, namely, the A-B and B-C interfaces, while the third cluster contains more variable, scenario-specific patterns.

In the first and second clusters, unmet demand in zone G is frequently associated with periods of high utilization and congestion at the A-B and B-C interfaces (Figure~\ref{fig:clus1_ABif_BCif_PJMif_utilization}a–b).
This suggests that significant pressure points arise along the entire upstate-to-downstate transmission corridor---crucially, also upstream of the well-known bottleneck at the E-G interface. 
Although the E–G interface is frequently or almost always congested across all scenarios (Figure~\ref{fig:scenario_features}c) and has been previously identified as a critical bottleneck \cite{liu_heterogeneous_2023,kabir_quantifying_2024}, its persistent congestion typically renders it non-predictive in this analysis (see Discussion).

The interface connecting NYS to the neighboring PJM system also emerges as a common pressure point in these clusters. 
In these scenarios, flow from NYS to PJM exhibits a strong diurnal pattern (Figure~\ref{fig:clus1_ABif_BCif_PJMif_utilization}c), increasing and decreasing with downstate demand.
This pattern reflects the NYS-PJM interface's important role in moving net power from upstate to downstate---critically, bypassing the bottleneck at the E-G interface---during periods of system stress.

The first and second clusters are primarily distinguished by the presence of the I-K interface as a predictive pressure point in the first cluster but not in the second.
Among the scenarios in which the I-K interface is identified as a predictive pressure point, we observe two distinct dynamics connecting I-K interface flow and power shortages in zone G.
In some scenarios, power shortages in zone G align with elevated interface flow in the I~$\to$~K direction, while elevated interface flow in the K~$\to$~I direction helps mitigate power shortages in zone G (Figure~\ref{fig:clus1_IKif_util}a). 
In other words, power moving out of zone K (during hours with excess offshore wind generation) helps the system meet demand in zone G, likely indirectly by balancing demand in zone J.

Other scenarios show persistent congestion and high interface flow in the K~$\to$~I direction, regardless of shortages in zone G (Figure~\ref{fig:s290_HIif_IKif_utilization}b). 
These scenarios are also characterized by increased flow from zone H to zone I, indicating overall higher downstate demand that draws heavily on external resources from other zones, both upstate zones and zone K (Figure~\ref{fig:s290_HIif_IKif_utilization}c). 
Under such conditions, exporting surplus electricity from zone K does not sufficiently reduce demand to prevent shortages in zone G; this is what we saw in the third, extreme temperature scenario examined previously.

To better understand what drives the emergence of these distinct pressure point patterns, we examine whether cluster membership can be predicted from high-level meteorological and technological features of each scenario. 
That is, we use a classification tree\cite{breiman_classification_2017} to examine to what extent broad weather and infrastructure conditions can account for the different types of reliability dynamics observed.
We consider five key variables in each scenario: the mean anomaly relative to a baseline scenario with historical meteorological patterns for four different metrics---(1) hourly temperature averaged across the state, (2) hourly solar availability, (3) hourly wind availability, and (4) quarter-monthly hydro availability---as well as the battery capacity scaling factor. 
In principle, these features reflect major drivers of electricity supply and demand and should help explain observed differences in predictive transmission pressure points.
However, a basic classification tree\cite{breiman_classification_2017} trained to predict cluster membership achieves only 51.4\% accuracy—just slightly better than random guessing (see Methods for details).

\begin{figure}[hbtp]
    \centering
    \includegraphics[width=3.5in]{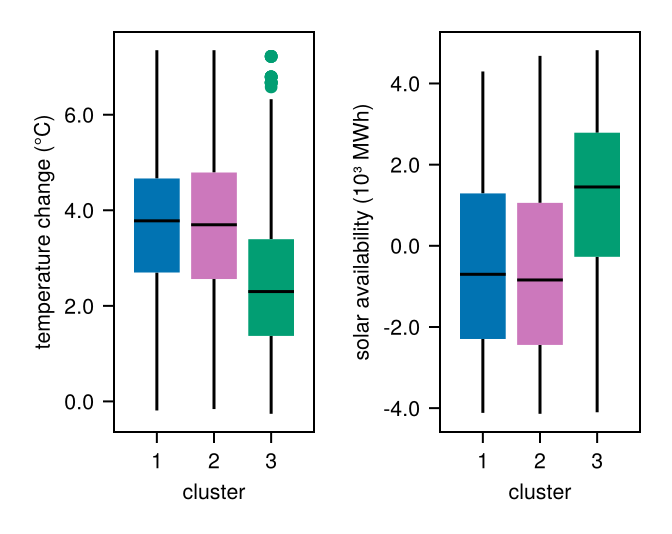}
    \caption{Distribution of mean hourly solar availability anomaly and mean hourly temperature anomaly relative to baseline scenario for three identified clusters.}
    \label{fig:cluster_solar_temp_boxplots}
\end{figure}

Still, examining the most influential predictors in the classification tree---the mean hourly anomaly in temperature and solar availability---reveals some broad connections between specific meteorlogical-technological features and reliability dynamics. 
Scenarios comprising the first and second cluster tend to have larger temperature increases---and therefore higher levels of electricity demand and lower availability of hydropower---combined with lower solar availability (Figure~\ref{fig:cluster_solar_temp_boxplots}). 
With simultaneously higher load concentrated downstate and relatively low solar availability, the grid is likely more reliant on upstate land-based wind and hydropower. 
This leads to pressure points at key upstate-to-downstate transmission interfaces---both moving power out of zones A and B and moving power through PJM in order to bypass the frequently congested E-G interface. 
However, the first and second clusters remain largely indistinguishable, even when considering the full set of features.

Scenarios belonging to the third cluster tend to have higher solar availability relative to the level of temperature increase.
As a result, these scenarios can be considered more well-behaved, and the sets of predictive pressure points that emerge are less systematic and more scenario-specific.

Overall, the limited predictive power of these features emphasize that transmission pressure points arise from complex interactions between generation, transmission, and demand across space and time. 
More granular metrics that better capture the spatiotemporal patterns of intermittent energy resources and electricity demand may offer better predictive insight. 
In the meantime, these results highlight the value of real-time grid monitoring and underscore the need for flexibility in reliability planning. 
Efforts to enhance grid reliability should remain adaptive, with strategies that can respond to evolving meteorological and technological conditions.

\section*{Discussion}
\label{sec:discussion}

Across a wide range of technological and meteorological conditions, our transfer entropy analysis consistently identifies common pressure points that reflect potentially persistent bottlenecks in a future NYS grid. 
In particular, predictive pressure points for downstate power shortages frequently emerge at interfaces all along the upstate-to-downstate transmission corridor, \emph{e.g.}, the A-B, B-C, and NY-PJM interfaces. 
Notably, these predictive pressure points are geographically and electrically distant from where the power shortages actually occur. 
This finding reinforces previous observations that the complex nature of the electric power system can result in a spatial separation between the affected generation or transmission infrastructure and where unmet demand is ultimately observed.\cite{webster_integrated_2022}

These predictive pressure points emerge from interactions between generation and transmission under system stress, particularly during periods of elevated demand---largely driven by higher temperatures---and limited energy availability.
Analyzing power flow dynamics at these transmission pressure points can be quite complex; for example, many power shortage events occur when transmission lines are not congested, which may seem counterintuitive. 
However, this reduced utilization can signal insufficient upstream generation, leaving the transmission paths underutilized. 
This adds nuance to a common diagnostic, the use of line congestion as a signal of a reliability risk.\cite{liu_heterogeneous_2023,kabir_quantifying_2024,webster_integrated_2022}

When power shortages primarily arise from reduced generation---as indicated by low curtailment but also low transmission utilization---expanding storage is likely to be a key reliability intervention. Storage can help balance the temporal mismatch between load and resource availability, particularly in solar-dominated systems.\cite{liu_heterogeneous_2023}
In these cases, expanding transmission would not directly address reliability challenges, though it might help move excess electric power through the system to storage facilities.

In other cases, such as the extreme temperature scenario, expanding storage capacity is not a straightforward or sufficient solution.
Severe transmission congestion leads to curtailment of wind and solar generation and downstream power shortages, necessitating either upgrades to transmission infrastructure in conjunction with expanded storage capacity or complementary, longer-range storage options such as green hydrogen, which could be transported using alternative infrastructure\cite{liu_heterogeneous_2023}.

Another possible solution is demand response. 
In this analysis, we choose to prioritize being able to meet all demand, so we do not explicitly consider electrical loads at other buses as potential pressure points.
Future work will study the role of demand-side stresses and interventions.

It is worth discussing some key pieces of infrastructure not identified as predictive pressure points by our pairwise transfer entropy analysis. Most notably, the E-G interface is identified as a predictive pressure point in less than 2\% of the full scenario ensemble despite being physically connected to our target load bus and having an obvious causal relationship with our ability to meet demand at this bus (Figure~\ref{fig:EGif_examples}).
This is the result of how transfer entropy is defined and illustrates an inherent limitation to using (pairwise) transfer entropy to define pressure points.

Transfer entropy measures predictive information \textit{gain}---that is, how much knowledge of E-G interface flow improves predictions of power shortages in zone G beyond what is already predictable from the history of unmet demand. 
When the E-G interface is persistently congested, as is the case in a large share of scenarios (Figure~\ref{fig:scenario_features}c), the time series reflecting interface utilization is very uninformative from a prediction standpoint: it is almost constant (top panel of Figure~\ref{fig:EGif_examples}b). 
This is analogous to the toy example presented in Figure~\ref{fig:basic_te_example}a, where the two processes are totally uncoupled.
On the other hand, when power shortages are largely the direct result of congestion in the E-G interface---\emph{i.e.}, the occurrence of power shortages is almost perfectly correlated with congestion---then the past history of interface flow does not provide any unique predictive information not already contained in the past history of unmet demand (bottom panel of Figure~\ref{fig:EGif_examples}b). 
This corresponds to the perfectly coupled case presented in Figure~\ref{fig:basic_te_example}c.

Another limitation of our analysis is that we have only considered the pairwise transfer entropy between supply or transmission infrastructure and a load bus.
Recovering the full set of interactions between components would require estimating multivariate transfer entropies.
Multivariate transfer entropy\cite{lizier_multivariate_2012,faes_information-based_2011} accounts for redundant interactions and detects synergistic (higher-order) interactions by conditioning on additional source processes, that is, by measuring the amount of information about target process $Y$ contained in past values of source process $X$, given knowledge of $Y$'s past \textit{and} the past of additional source processes $Z_1,\dots,Z_n$.
We can see an example of a false positive identification due to redundancy in the well-behaved scenario: solar availability upstream, particularly in zone F, drives both power shortages in zone G and the dynamics of flow through the G-H interface. 
This common driver creates a false predictive relationship from G-H interface flow to power shortages in zone G.

Conversely, our pairwise analysis almost surely overlooks some synergistic interactions---most notably, the role of batteries in helping compensate for transmission congestion and intermittent energy resources.
Across the three narrative scenarios, no batteries are ever identified as predictive pressure points, likely because their own dynamics are closely tied to transmission congestion.
Discharging batteries upstream of interface congestion cannot help relieve power shortages in zone G, and depleted batteries downstream of transmission congestion cannot be recharged. 
Estimating a multivariate transfer entropy from the pattern of battery utilization to the pattern of power shortages with relevant interface congestion taken into account could help identify this synergistic interaction between storage and interface flow as a (predictive) pressure point in the system.

A brute force method for full network inference via multivariate transfer entropy would likely be computationally intractable given the combinatorial explosion in the number of potential sets of candidate sources.\cite{lizier_multivariate_2012} 
Thus, methods for full network inference via multivariate transfer entropy\cite{lizier_multivariate_2012, faes_information-based_2011,novelli_large-scale_2019} typically build the conditioning set of relevant source processes iteratively.
However, these methods are still computationally intensive, requiring many transfer entropy estimates and tests for statistical significance (see Methods).
In future work, it is worth considering how domain knowledge and knowledge of the topology of the power grid may be helpful in expanding the set of relevant source processes. 
It is also worth exploring whether other methods of reconstructing (higher-order) networks from time series data\cite{becker_large-scale_2023, casadiego_model-free_2017} may be more appropriate.

Despite these limitations, this analysis has demonstrated that transfer entropy can be a useful tool in identifying infrastructure pressure points emerging as a result of complex interactions between generation and transmission patterns in a grid weakened by natural stressors.
Addressing these challenges to reliability often requires investment in or upgrades to multiple types of infrastructure---energy generation, transmission, and storage---in tandem.
However, the emergence of specific pressure points is difficult to anticipate from high-level grid operating conditions alone, underscoring the need for adaptive planning frameworks and real-time monitoring tools that can respond to emergent risks rather than relying solely on predictive modeling.

\section*{Methods}
\label{sec:methods}

\subsection*{Model \& scenario design}

We use a scenario ensemble generated by ACORN\cite{liu_heterogeneous_2023}, a reduced-form DC optimal power flow (DC-OPF) model of the proposed 2040 NYS power grid outlined in the CLCPA scoping plan\cite{new_york_state_climate_action_council_new_2022}.
The scoped plan includes (relative to the present-day NYS grid~\cite{liu_open_2022}) expanded variable energy generation and transmission capacity, new HVDC lines, and additional load due to electrification of transportation and heating/cooling sectors.

As grid reliability is heavily influenced by the  joint behavior of temperature (which affects demand and transmission ampacity) and energy supply (solar, wind, and hydro), it is important for operational assessments to use realistic joint distributions of these meteorological inputs. 
To maintain their co-variability, we force the simulations using reanalysis weather data from MERRA2 over the 22-year simulation period (1998 to 2019).
The DC-OPF problem is solved at an hourly resolution to determine the optimal dispatch of resources that minimizes power shortages while ensuring operational constraints (\emph{e.g.}, transmission capacity and phase angle constraints, quarter-monthly hydro requirements, battery state transitions) are enforced. 
Due to the lack of cost-setting generators in the proposed 2040 NYS grid, ACORN does not minimize cost of operating the system but rather maximizes reliability.

In order to account for deep uncertainties\cite{srikrishnan_uncertainty_2022} surrounding natural meteorological variability and technological changes (such as temperature changes, variable energy integration, battery storage expansion, and electrification) we employ the full set of 300 alternative scenarios from Liu et al.\cite{liu_heterogeneous_2023}
Using temperature-adjusted meteorological inputs in the model's supply, demand, and transmission modules, the authors construct scenario-specific time series for the availability of intermittent energy resources, electrified demand, and transmission line ampacity.
See Liu et al.\cite{liu_heterogeneous_2023} for further details.

We consider each year of the 22-year simulation period for each alternative meteorological-technological scenario as a different scenario-year, giving us an ensemble of 6,600 scenario-years.
Meteorological and technological features for each scenario-year are summarized by anomalies relative to a baseline scenario averaged across the 22-year simulation period: we compute mean hourly anomaly in time series of statewide (1) average temperature, (2) solar availability, (3) wind availability, and (4) load as well as the mean quarter-monthly anomaly for statewide hydro availability.
The baseline scenario incorporates the changes in generation technology, transmission infrastructure, storage expansion, and load profile detailed in the CLCPA scoping plan\cite{new_york_state_climate_action_council_new_2022} but only uses historical meteorological patterns.
The final technological feature of each scenario-year is the battery capacity scaling factor, which ranges from 0.6 to 1.4.

\subsection*{Transfer entropy estimation \& significance testing}

Transfer entropy (TE) is a directed (\emph{i.e.}, asymmetric) measure of predictive information transfer from a source process $X$ to a target process $Y$. 
Specifically, TE from $X$ to $Y$ is the mutual information between the previous state variable $X^-$ of source $X$ and the future value $Y^+$ of target $Y$, conditioned on the past state variable $Y^-$ of target $Y$:
\begin{align*}
    TE(X\to Y) &= I(X^- : Y^+ \mid Y^-)\\
    &= \sum_{x^-,\ y^+,\ y^-} p(x^-, y^+, y^-)\log\frac{p(y^+ \mid y^-, x^-)}{p(y^+ \mid y^-)}
\end{align*}
This definition\cite{bossomaier_introduction_2016} assumes $X$ and $Y$ are discrete-valued stationary processes; there is an analogous definition for continuous-valued processes\cite{kaiser_information_2002} based on differential entropy.

By conditioning on the past, transfer entropy incorporates directional and dynamical information; however, it is a measure only of observed correlation rather than causal effect, which typically requires some perturbation of or intervention in the system to measure.\cite{bossomaier_introduction_2016}
Transfer entropy is therefore appropriately interpreted as a model-free test statistic for Granger causality, also known as observational causality.\cite{wibral_transfer_2014}

When estimating TE from time-series data, it is convenient if $X$ and $Y$ can be assumed to be stationary processes so that the appropriate probability distributions involving $X^-$, $Y^-$, and $Y^+$ can be estimated from samples over time.\cite{wibral_transfer_2014,bossomaier_introduction_2016}
(See Wollstadt et al.\cite{wollstadt_efficient_2014} and Gómez-Herrero et al.\cite{gomez-herrero_assessing_2015} for details on estimating time-dependent TE values from ensembles of time series representing copies of the random processes.)
Thus, to limit the influence of seasonal and subseasonal trends that could introduce nonstationarity, we restrict our analysis to the summer months---June, July, and August---and scale the time series by relevant capacity as described below.

For each potential predictive pressure point $X$, we follow a standard procedure\cite{wibral_transfer_2014,lindner_trentool_2011} for estimating pairwise TE from the time series $\{x_t\}_{t=1}^T$ measuring utilization of the resource or infrastructure and the time series $\{y_t\}_{t=1}^T$ measuring power shortages at the target load bus.
We reconstruct the underlying state space for each process by assuming that $Y^-$ and $X^-$ take the form of time-delay embedding vectors\cite{wibral_transfer_2014,wollstadt_efficient_2014,lindner_trentool_2011} and optimizing the embedding parameters via the Ragwitz criterion\cite{ragwitz_markov_2002}. 
It is standard to take $Y^+$ to be the value of target $Y$ one time-step in the future.
This state space reconstruction step is critical to avoid under- or overestimating $TE(X\to Y).$
See SI for more details on state space reconstruction.

We estimate $TE(X\to\ Y)$ using the Associations.jl Julia package\cite{haaga_associations_2025}.
After scaling, the time series for resources that predominantly jump between no utilization and full utilization become sufficiently discrete; in this case, estimating the relevant probabilities in the TE functional by the frequency of occurrence of different states\cite{wibral_transfer_2014} provides the most consistent results.
For continuous-valued processes, state-of-the-art TE estimators use nearest neighbor estimates of the relevant probability densities with adaptive resolutions to account for the different dimensionalities of the underlying densities.\cite{kraskov_estimating_2004}
We use the FPVP estimator\cite{frenzel_partial_2007,vejmelka_inferring_2008} in this case.

Raw transfer entropy estimates have limited interpretability because of inherent statistical bias in estimators and because of the varying statistical properties of the underlying time series.\cite{bossomaier_introduction_2016} 
Thus, to determine when a transfer entropy estimate suggests meaningful information transfer, we rely on a null-hypothesis significance test, estimating the distribution of $TE(X\to Y)$ under the null hypothesis that there is no predictive information transfer from $X$ to $Y$ and then calculating the $p$-value for obtaining the sample estimate of $TE(X\to Y)$ under this null distribution.\cite{bossomaier_introduction_2016,wibral_transfer_2014}
We estimate this null distribution by generating surrogate source time series\cite{keylock_constrained_2006} $\{\hat{x}\}_{t=1}^T$ that preserve important statistical properties of the original source time series $\{x\}_{t=1}^T$ and use a significance threshold of 0.05. (See SI for more details on surrogate time series generation.)
Lowering the significance level reduces the number of predictive pressure points identified in each scenario-year, decreasing the likelihood of false positives but potentially overlooking meaningful interactions.
Conversely, increasing the significance level makes the test more permissive, increasing the chance of detecting weaker but possibly spurious correlations.

\subsection*{Simulation data for transfer entropy estimation}

Hourly time series reflecting curtailment, generation, interface flow, and battery state are used in transfer entropy estimates---representing source processes---to identify pressure points at wind and solar generators, hydro generators, transmission interfaces, and batteries, respectively.
We consider curtailment as a metric for non-dispatchable energy resource utilization because it is more straightforward to interpret than generation. 
A lack of generation could reflect either curtailment or unavailability of that resource, while a lack of curtailment indicates that we are fully utilizing that resource's available energy.
We consider interface flow across the NYS zonal interfaces to capture transmission utilization instead of individual branch flow because it gives a more comprehensive picture of how power is moving through the grid. 
Flows through individual branches are noisier and may be highly sensitive to small variations in DC-OPF model parameters. 

We scale each time series by the relevant capacity---hourly energy availability and transmission capacity for intermittent energy resources and transmission interfaces, respectively, and installed capacity for hydropower generation and batteries---to reflect when resources are being under- or fully utilized.
This captures the weakening of the system due to weather-related stressors.
As an example, when extreme temperatures decrease the carrying ampacity of transmission lines, the amount of electricity which can flow through the lines is limited.\cite{bartos_impacts_2016}
Any resulting shortage could be associated with the ``true" congestion rather than the nominal under-utilization.
Rescaling to unit range is also important to ensure accurate and stable transfer entropy estimates.

The target process in all of our transfer entropy estimates is hourly unmet demand (as a proportion of total demand) at a selected zone G bus.
This bus was chosen as it consistently experiences the most frequent occurrence and highest proportion of unmet demand across all scenario-years (Figure~\ref{fig:ls_hrs_and_prop_by_bus}).

\subsection*{Scenario discovery}

To identify meteorological and technological features that result in consistent patterns of predictive transmission pressure points, we cluster the results of our transfer entropy analysis using the $k$-modes algorithm, a variation of the $k$-means algorithm suitable for categorical data.\cite{ng_impact_2007,huang_extensions_1998} 
For each scenario-year, the flow at each of the 15 transmission interfaces either has or does not have significant transfer entropy to the pattern of power shortages at the zone G load bus.

Using a common approach to determine the number of clusters---examining the within-cluster dissimilarity---suggests there could be up to eight distinct clusters (Figures~\ref{fig:cluster_obj_vs_k} and \ref{fig:map_8_clusters}).
However, there is considerable overlap across these clusters: for example, four of the largest clusters represent different combinations of predictive pressure points at the A-B, B-C, NYS-PJM, and I-K interfaces.
Furthermore, the smaller clusters are considerably less cohesive, suggesting that there is a fraction of scenario-years with atypical combinations of predictive transmission pressure points.
Thus, we choose to limit our analysis to three clusters (see Figure~\ref{fig:map_3_clusters}) representing the most dominant patterns.

We use a classification tree\cite{breiman_classification_2017} trained using the MLJ.jl Julia package\cite{blaom_mlj_2020} to predict cluster membership using a subset of the meteorological and technological features described above: mean hourly anomaly in temperature, solar availability, wind availability, and hydro availability as well as battery capacity scaling factor. 
We exclude mean hourly load anomaly as this is almost perfectly correlated with the mean hourly anomaly in temperature (Figure~\ref{fig:load_and_hydro_vs_temp}b).
To avoid potential overfitting, we limit the depth of the tree to three levels.
The fitted classification tree is in Figure~\ref{fig:class_tree}. 
As discussed in the results, this tree is a relatively weak classifier; the balanced accuracy\cite{mosley_balanced_2013} is 51.4\%, which represents a fairly small improvement over random guessing. 
A classification tree for predicting membership among the full eight clusters is also a weak classifier; the balanced accuracy for a tree with four levels is 30.7\%---again, a small improvement over random guessing.

\subsection*{Data and code availability}

All code and a portion of the data used for this analysis is archived at \url{https://doi.org/10.5281/zenodo.16783066}. Any current unreleased code is available at \url{https://github.com/kbtang28/tang-etal_inprep}. 

\subsection*{Acknowledgments}

K.T. and V.S. were partially supported for this research by the U.S. Department of Energy, Office of Science, Biological and Environmental Research Program, Earth and Environmental Systems Modeling, MultiSector Dynamics under Cooperative Agreement DE-SC0022141. This research was also conducted with support from the Cornell University Center for Advanced Computing, which receives funding from Cornell University, the National Science Foundation, and members of its Partner Program. 

\subsection*{Author contributions}
K.T., conceptualization, methodology, formal analysis, visualization, writing - original draft; M.V.L., data curation, writing - reviewing \& editing; C.L.A., supervision, writing - reviewing \& editing; V.S., conceptualization, supervision, writing - reviewing \& editing.

\bibliographystyle{unsrt}
\bibliography{refs}

\newpage
\renewcommand{\thefigure}{S\arabic{figure}}
\setcounter{figure}{0}
\renewcommand{\thetable}{S\arabic{table}}

\section*{Supplementary material}

\begin{figure}[hbtp]
    \centering
    \makebox[\textwidth][c]{
        \includegraphics[height=2.75in]{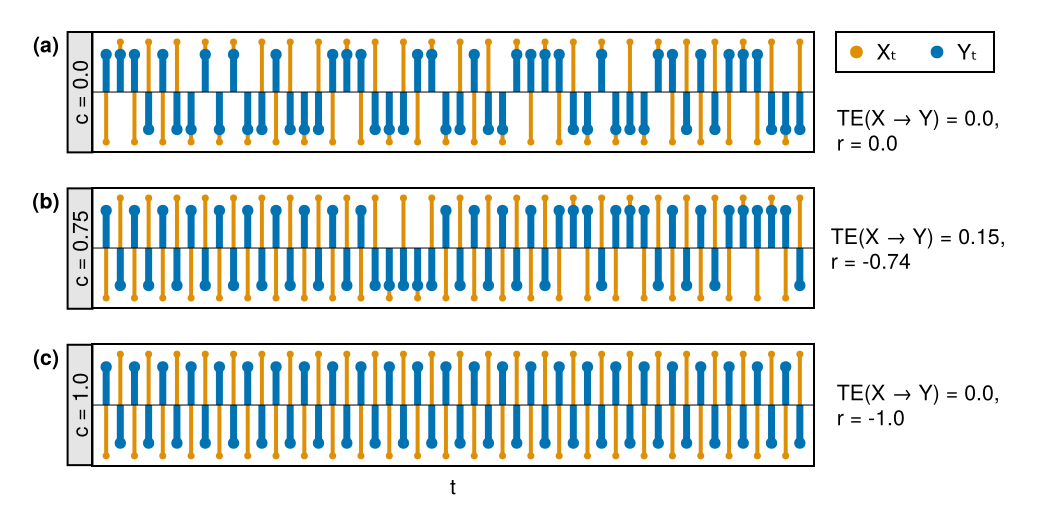}
    }
    \caption{Transfer entropy and correlation in a bivariate, first-order, stationary Markov chain. The process $X$ evolves deterministically, alternating between states $x$ and $-x$ at each timestep. The current state of $Y$ depends probabilistically on the last state of $X$; at time $t,$ $Y_t=X_{t-1}$ with probability $(1+c)/2$ and $Y_t=-X_{t-1}$ with probability $(1-c)/2,$ where $-1\leq c\leq 1.$ (Note $X_t$ and $Y_t$ are plotted with different scales above.) For each coupling strength, transfer entropy $TE(X\to Y)$ is computed analytically (see Example 4.1 in Bossomaier et al.\cite{bossomaier_introduction_2016}) and the Pearson correlation coefficient is estimated from 1000 simulated timesteps.}
    \label{fig:basic_te_example}
\end{figure}

\begin{figure}[hbtp]
    \centering
    \includegraphics[width=4.0in]{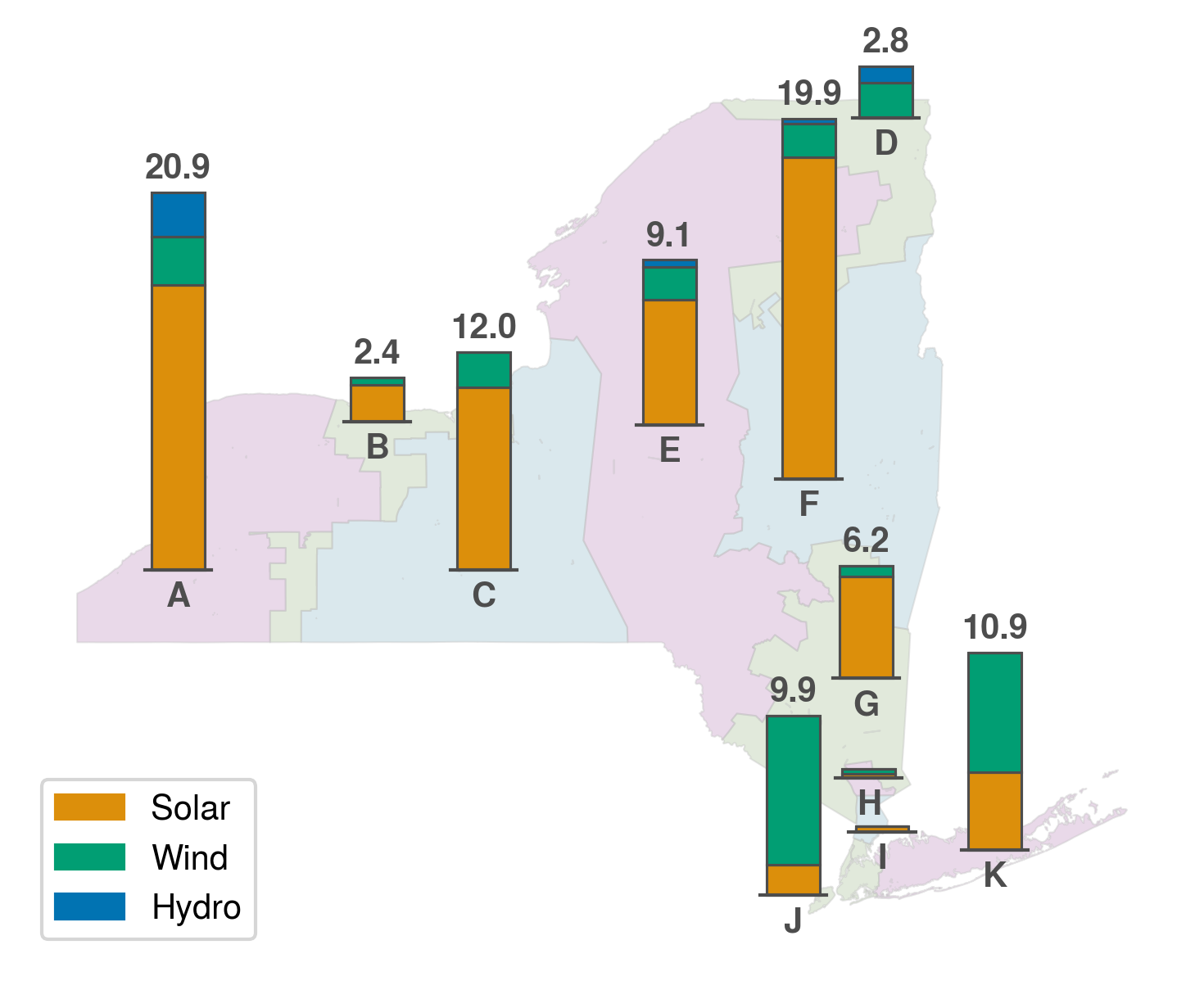}
    \caption{Planned solar, wind, and hydro resources by zone under the CLCPA\cite{new_york_state_climate_action_council_new_2022}. Total planned capacity is labeled for most zones; Zones H and I have a planned capacity of 505 MW and 299 MW, respectively.}
    \label{fig:map_zonal_caps}
\end{figure}

\begin{figure}[hbtp]
    \centering
    \includegraphics[width=4.0in]{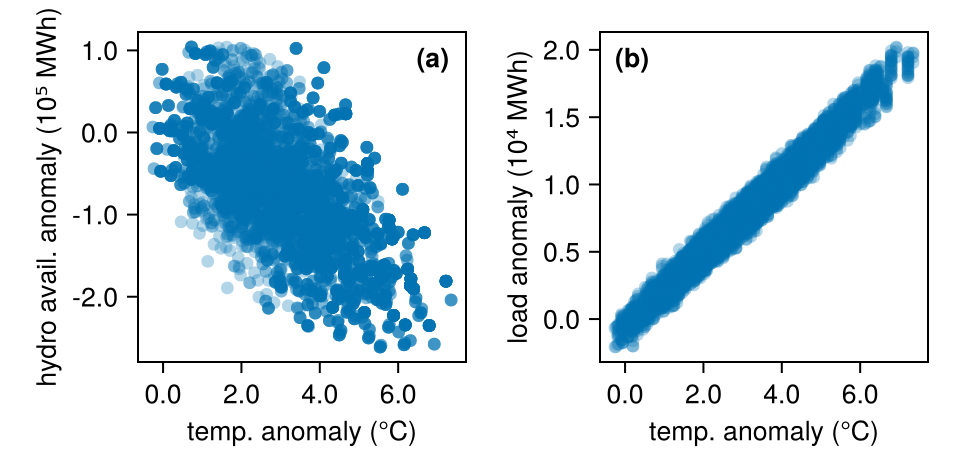}
    \caption{Mean anomaly in statewide (a) load and (b) hydro availability vs. mean temperature anomaly across the full scenario-year ensemble. These anomalies are measured relative to a baseline scenario---see Methods for details.}
    \label{fig:load_and_hydro_vs_temp}
\end{figure}

\begin{table}[hbtp]
\begin{center}
\renewcommand{\arraystretch}{1.2}
\label{table:scenario}
\footnotesize{
    \begin{tabular}{l|c|c|c|}
    \cline{2-4}
                                                                                                                                     & \textbf{Well-behaved}        & \textbf{\begin{tabular}[c]{@{}l@{}}Limited\\ resources\end{tabular}} & \textbf{\begin{tabular}[c]{@{}l@{}}Extreme\\ temperature\end{tabular}} \\ \hline
    \multicolumn{1}{|l|}{\textbf{\begin{tabular}[c]{@{}l@{}}Average temperature \\ anomaly ($\degree$C)\end{tabular}}}                 & 2.17 (0.26)              & 2.29 (0.28)                                                          & 5.34 (0.90)                                                            \\ \hline
    \multicolumn{1}{|l|}{\textbf{\begin{tabular}[c]{@{}l@{}}Average solar\\anomaly (MWh)\end{tabular}}}                   & $3.7\times 10^3$ (0.97)  & $-3.7\times 10^3$ (0.02)                                             & $4.2\times 10^3$ (0.99)                                                \\ \hline
    \multicolumn{1}{|l|}{\textbf{\begin{tabular}[c]{@{}l@{}}Average wind\\anomaly (MWh)\end{tabular}}}                    & $1.3\times 10^3$ (0.72)  & $-2.4\times 10^3$ (0.11)                                             & $-2.2\times 10^3$ (0.13)                                               \\ \hline
    \multicolumn{1}{|l|}{\textbf{\begin{tabular}[c]{@{}l@{}}Average hydro\\anomaly (MWh)\end{tabular}}}                   & $-6.0\times 10^4$ (0.59) & $-2.0\times 10^4$ (0.80)                                              & $-1.5\times 10^5$ (0.20)                                                \\ \hline
    \multicolumn{1}{|l|}{\textbf{\begin{tabular}[c]{@{}l@{}}Prop. of demand unmet\\at zone G load bus\end{tabular}}} & 0.38 (0.22)              & 0.68 (0.81)                                                          & 0.73 (0.88)                                                            \\ \hline
    \multicolumn{1}{|l|}{\textbf{\begin{tabular}[c]{@{}l@{}}Prop. of wind \& solar\\ energy curtailed\end{tabular}}}    & 0.42 (0.92)              & 0.07 (0.04)                                                          & 0.37 (0.81)                                                            \\ \hline
    \multicolumn{1}{|l|}{\textbf{\begin{tabular}[c]{@{}l@{}}Prop. of hours with\\IF E-G congested\end{tabular}}}              & 0.65 (0.04)              & 0.98 (0.94)                                                          & 0.97 (0.85)                                                            \\ \hline
    \end{tabular}
}
    \caption{Features of three chosen scenarios. Percentile for each value is in parentheses. Hourly temperature and generation capacity anomalies are measured relative to a baseline scenario---see Methods for details.}
\end{center}
\end{table}

\begin{figure}[hbtp]
    \centering
    \makebox[\textwidth][c]{
        \includegraphics[height=3.0in]{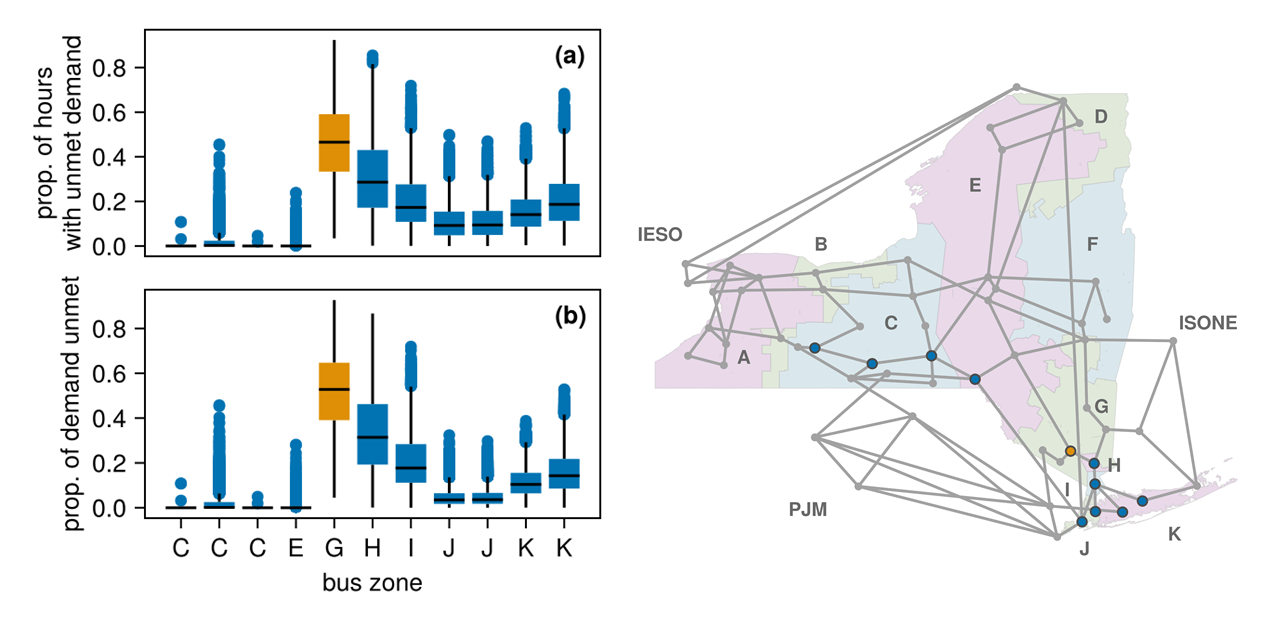}
    }
    \caption{Distribution of (a) the proportion of hours with unmet demand and (b) the proportion of total demand unmet across the full scenario-year ensemble at each bus experiencing power shortages. The zone G target bus used throughout our analysis is highlighted in yellow.}
    \label{fig:ls_hrs_and_prop_by_bus}
\end{figure}

\begin{figure}[hbtp]
    \centering
    \makebox[\textwidth][c]{
        \includegraphics[height=6.0in]{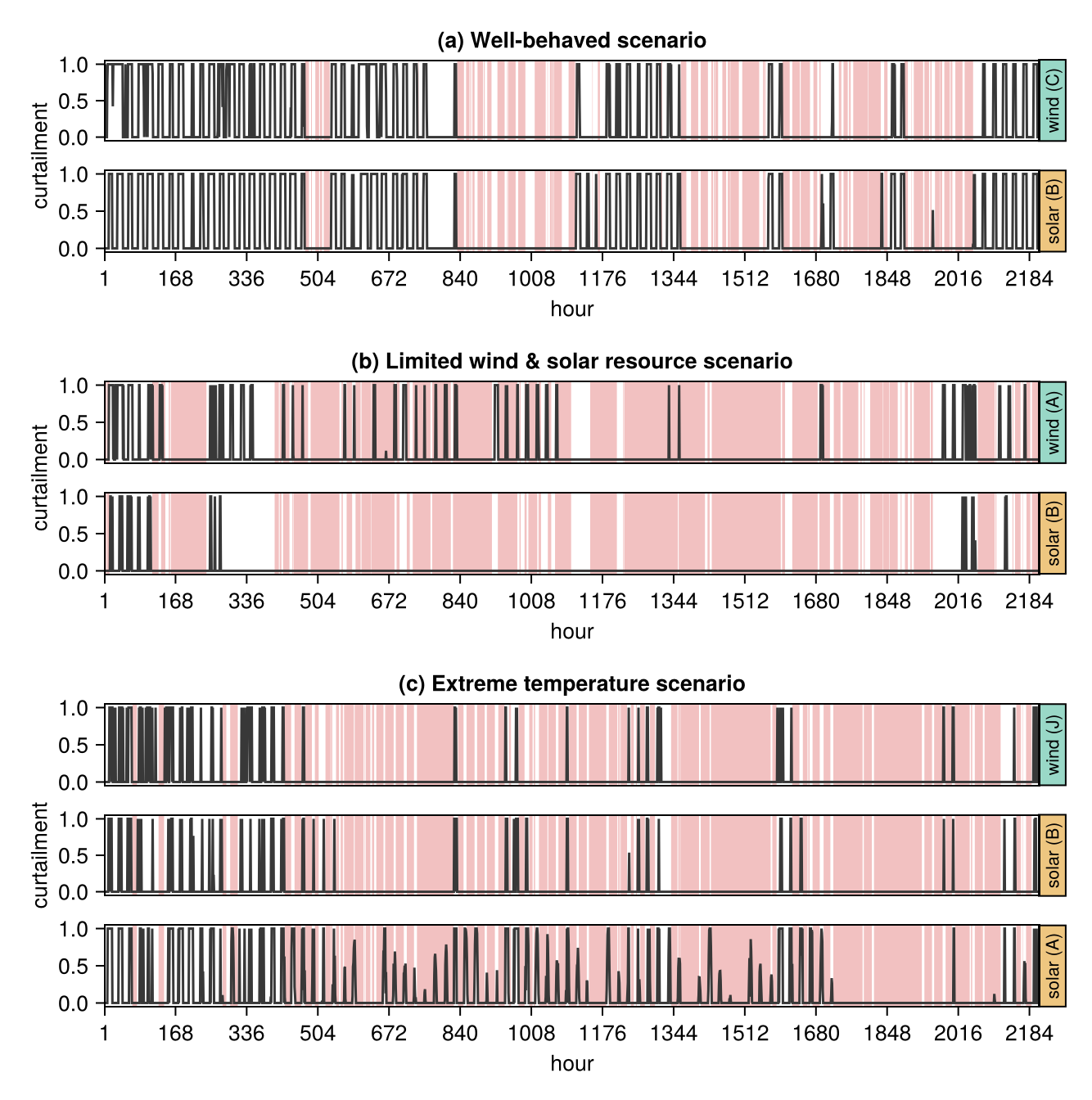}
    }
    \caption{Representative curtailment time series for individual predictive supply-side pressure points in the (a) well-behaved, (b) limited wind \& solar, and (c) extreme temperature scenarios. Curtailment is scaled by available wind or solar generation capacity in each hour, \emph{i.e.}, 0.0 indicates full utilization of the available resource and 1.0 indicates full curtailment. Hours with power shortages at the zone G target bus are shaded in red. Note that in the extreme temperature scenario, we see curtailment during hours with unmet demand at two predictive pressure points, one of which is shown in the bottom panel of (c).}
    \label{fig:curtailment}
\end{figure}

\begin{figure}[hbtp]
    \centering
    \makebox[\textwidth][c]{
        \includegraphics[height=2.0in]{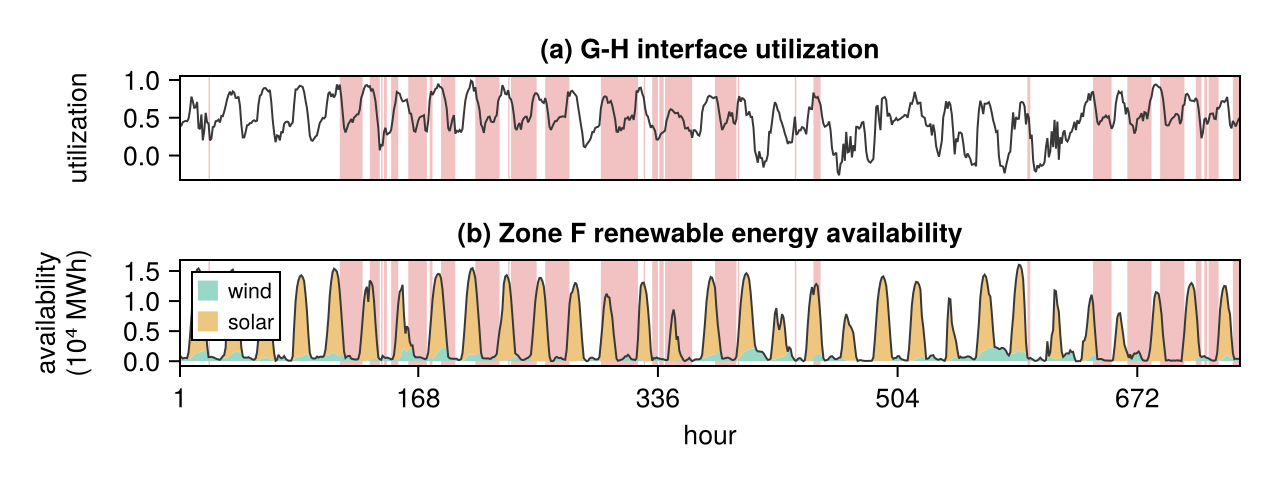}
    }
    \caption{(a) Utilization of the G-H interface and (b) availability of wind and solar energies in zone F over the course of one month in the well-behaved scenario. Hours with power shortages at the zone G target bus are shaded in red.}
    \label{fig:s140_GHif_zoneFrenewables}
\end{figure}

\begin{figure}[hbtp]
    \centering
    \makebox[\textwidth][c]{
        \includegraphics[height=2.75in]{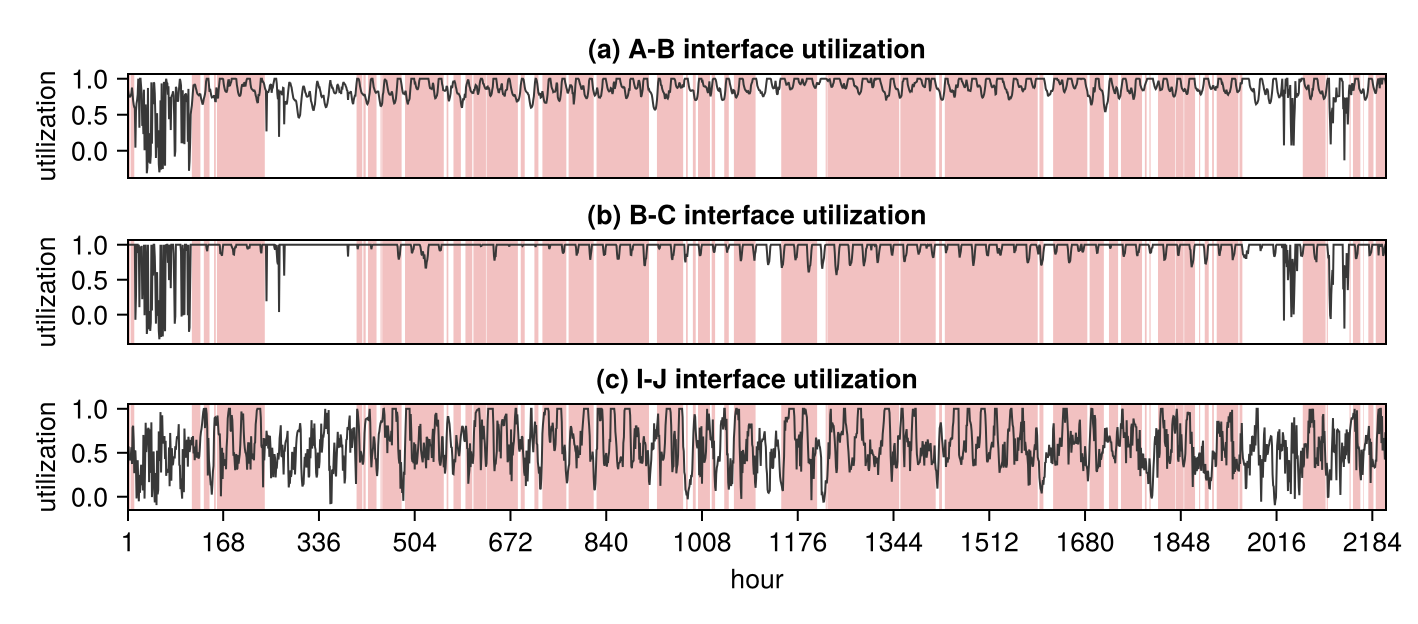}
    }
    \caption{Utilization of the (a) A-B, (b) B-C, and (c) I-J interfaces in the limited wind and solar resource scenario. Hours with power shortages at the zone G target bus are shaded in red.}
    \label{fig:s69_ABif_BCif_IJif_utilization}
\end{figure}

\begin{figure}[hbtp]
    \centering
    \makebox[\textwidth][c]{
        \includegraphics[height=3.5in]{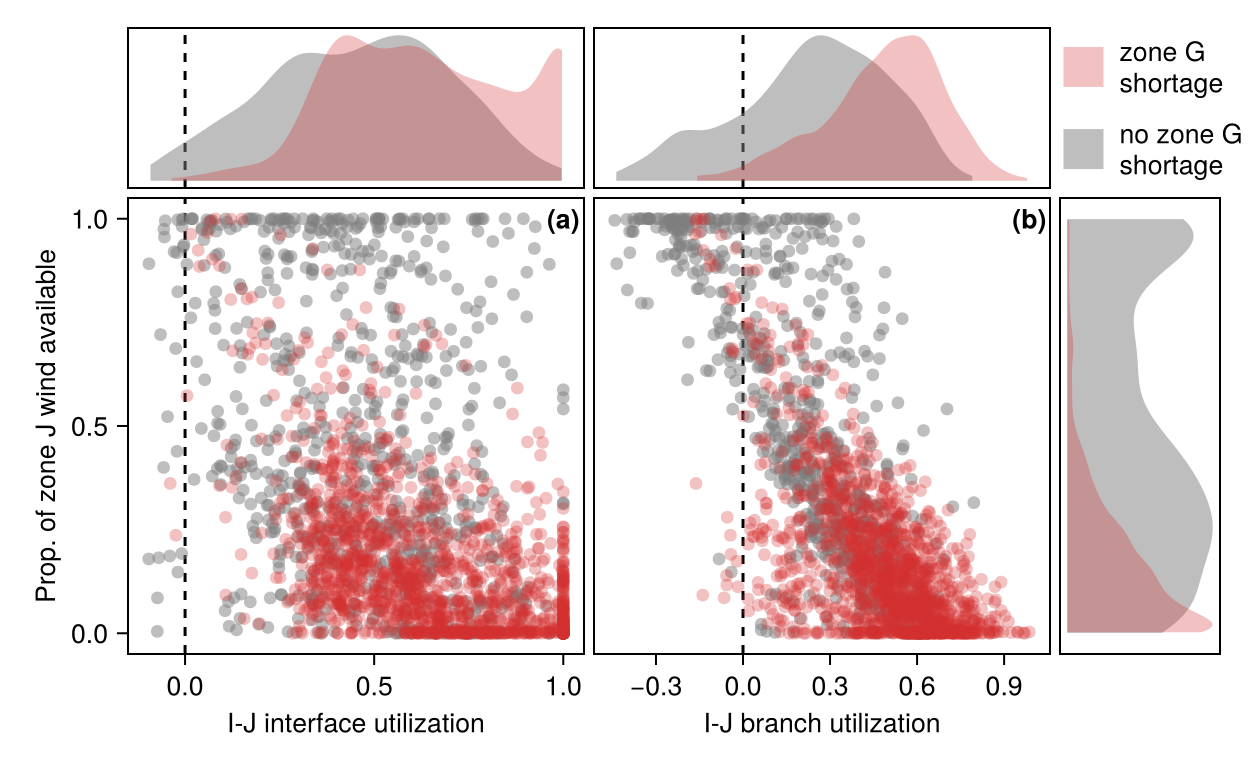}
    }
    \caption{Proportion of zone J's offshore wind capacity available vs. utilization of (a) the I-J interface and (b) one branch in the I-J interface for each hour of the simulation period in the limited wind and solar resource scenario. Positive and negative utilization values represent flow in the I~$\to$~J and J~$\to$~I direction, respectively. Note that there are two branches involved in the I-J interface; through the other branch, power always flows in the I~$\to$~J direction.}
    \label{fig:s69_zoneJwind_IJif_scatter}
\end{figure}

\begin{figure}[hbtp]
    \centering
    \makebox[\textwidth][c]{
        \includegraphics[height=4.5in]{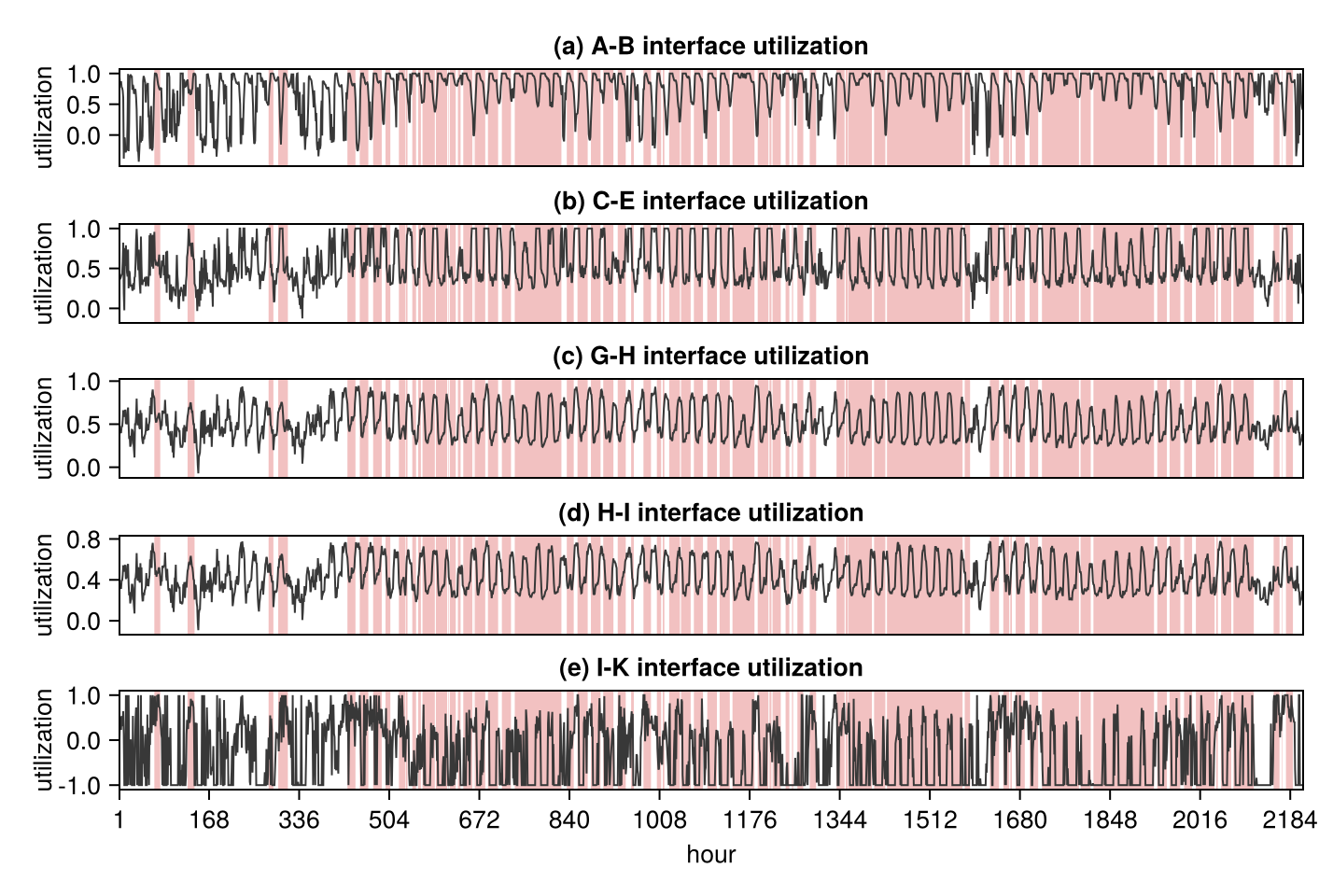}
    }
    \caption{Utilization of the (a) A-B, (b) C-E, (c) G-H,  (d) H-I, and (e) J-K interfaces in the extreme temperature scenario. Hours with power shortages at the zone G target bus are shaded in red.}
    \label{fig:s290_ABif_CEif_GHif_HIif_IKif_utilization}
\end{figure}

\begin{figure}[hbtp]
    \centering
    \makebox[\textwidth][c]{
        \includegraphics[height=2.25in]{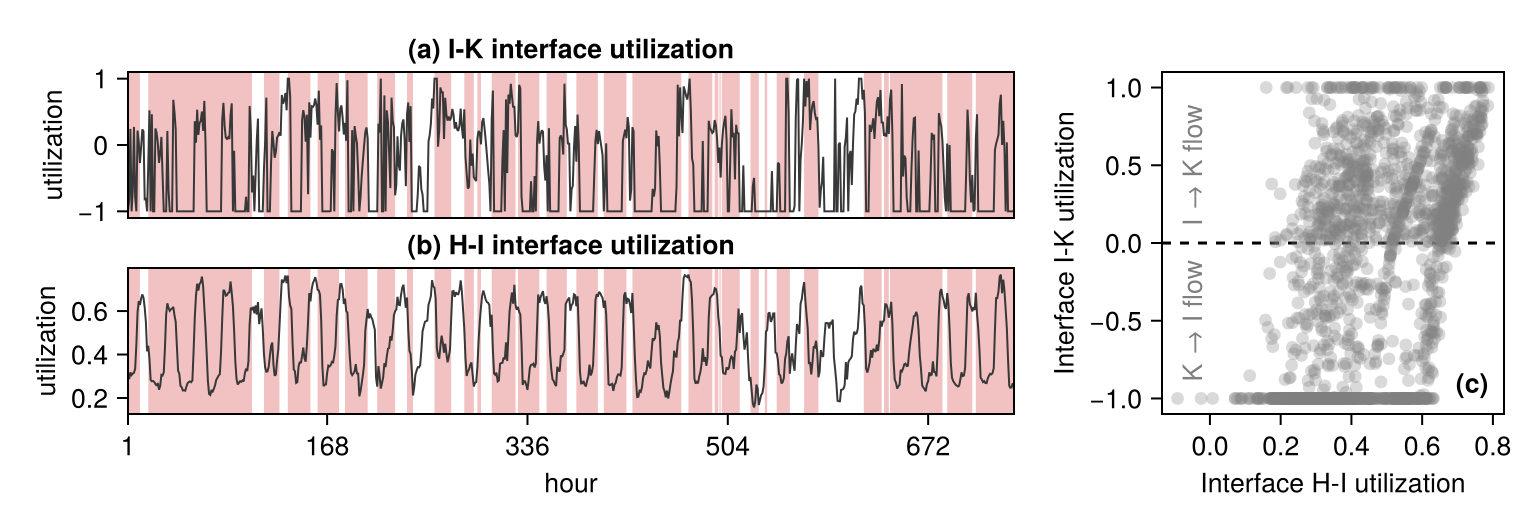}
    }
    \caption{Utilization of the (a) I-K and (b) H-I interface over one month in the extreme temperature scenario. Hours with power shortages at the zone G target bus are shaded in red. The utilization of the I-K vs. H-I interface is plotted for each hour of the simulation period in (c).}
    \label{fig:s290_HIif_IKif_utilization}
\end{figure}

\begin{figure}[hbtp]
    \centering
    \makebox[\textwidth][c]{
        \includegraphics[height=2.75in]{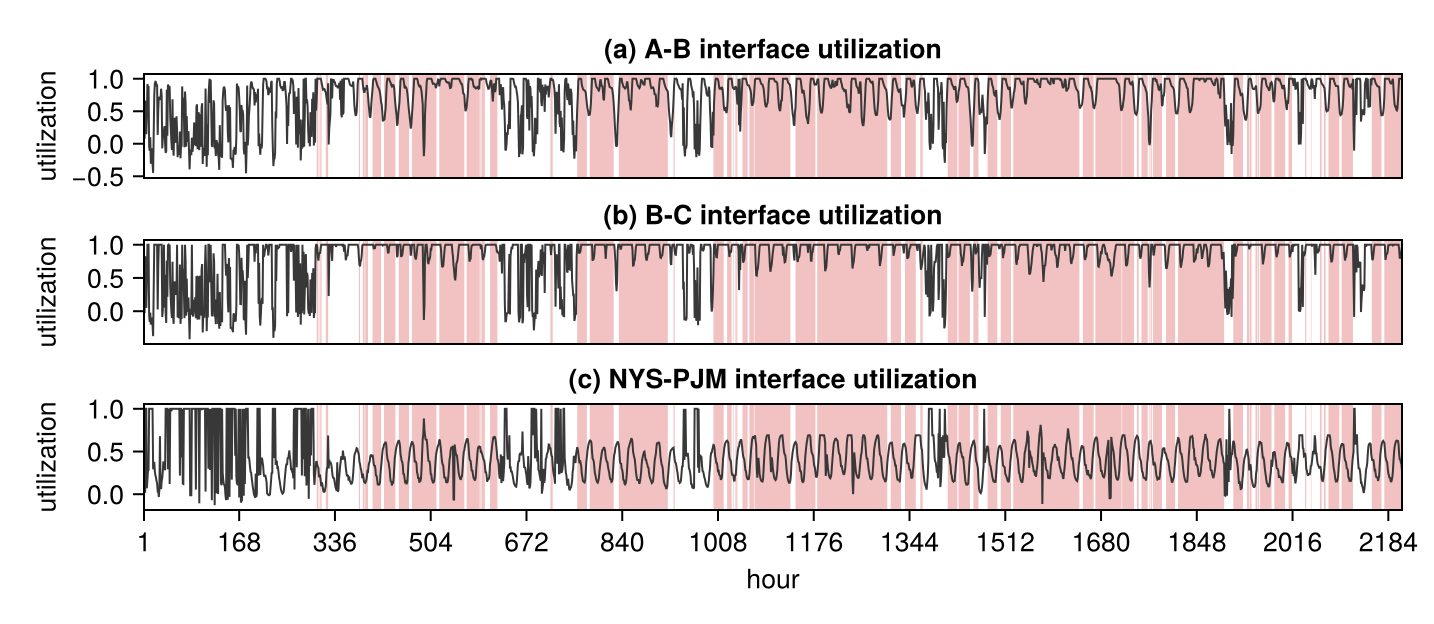}
    }
    \caption{Utilization of the (a) A-B, (b) B-C, and (c) NYS-PJM interface for a representative scenario belonging to Cluster 1. Hours with power shortages at the zone G target bus are shaded in red. In this scenario, all three of these interfaces are identified as predictive transmission pressure points.}
    \label{fig:clus1_ABif_BCif_PJMif_utilization}
\end{figure}

\begin{figure}[hbtp]
    \centering
    \makebox[\textwidth][c]{
        \includegraphics[height=2.5in]{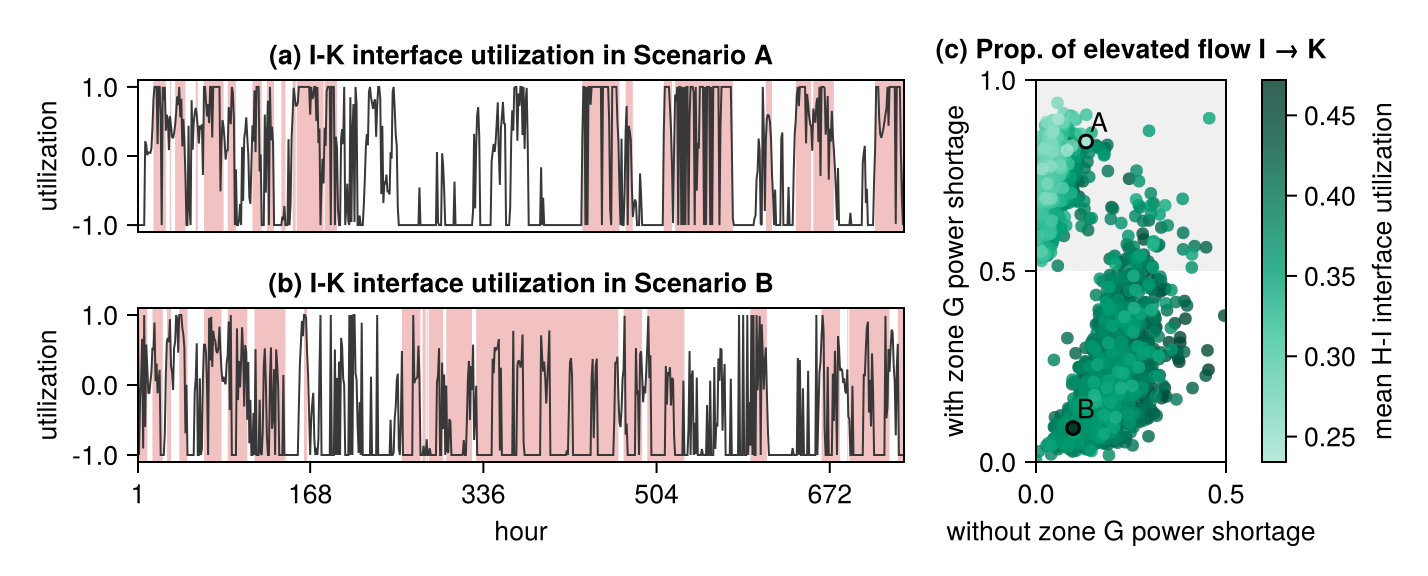}
    }
    \caption{Two distinct dynamics between I-K interface utilization and power shortages at the target zone G bus in scenarios belonging to cluster 1. Utilization of the I-K interface over one month is shown for two representative scenarios in (a) and (b). Hours with power shortages at the zone G target bus are shaded in red. In (c), the proportion of hours in which elevated flow (utilization of $\leq -0.9$ or $\geq$ 0.9) occurs in the I~$\to$~K direction is plotted for each scenario belonging to cluster 1. That is, 0.0 indicates that all hours with elevated flow are in the K~$\to$~I direction, and 1.0 indicates that all hours with elevated flow are in the I~$\to$~K direction. Notably, without power shortages in zone G there is never a majority of elevated flow in the I~$\to$~K direction. Higher mean utilization of the H-I interface can be interpreted as a proxy indicator for overall higher downstate demand that draws heavily on external resources from other zones.}
    \label{fig:clus1_IKif_util}
\end{figure}

\begin{figure}[hbtp]
    \centering
    \makebox[\textwidth][c]{
        \includegraphics[height=2.75in]{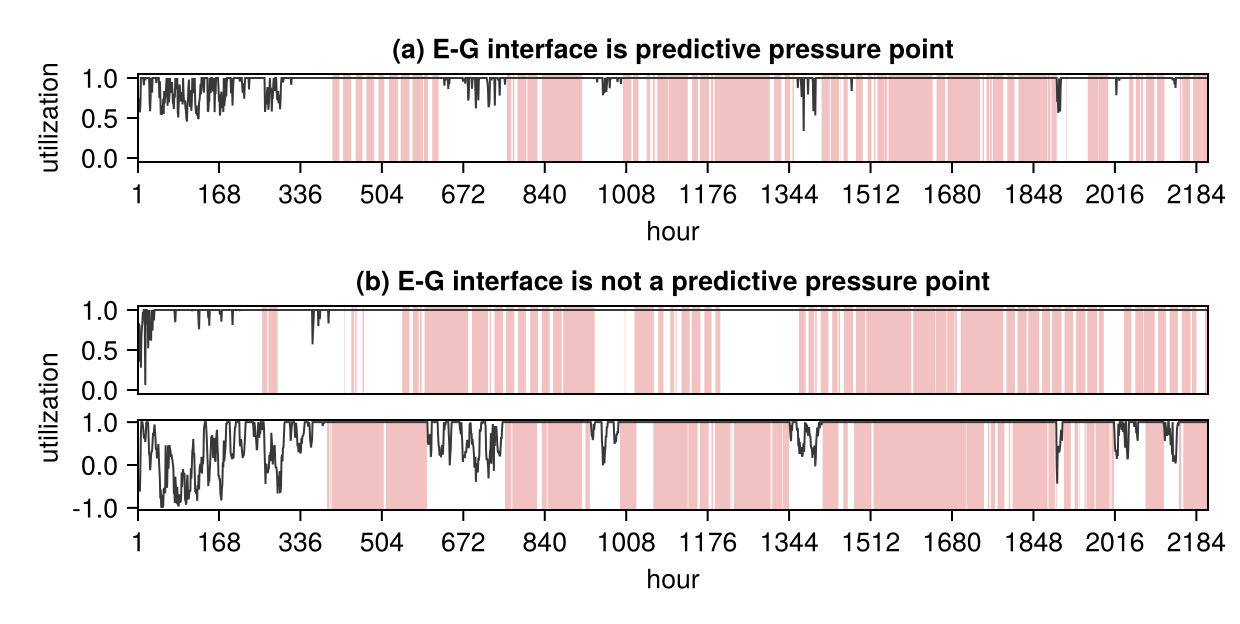}
    }
    \caption{Examples of E-G interface utilization. A scenario in which the E-G interface is identified as a predictive transmission pressure point is shown in (a); two scenarios in which the E-G interface is not identified as a predictive transmission pressure point are shown in (b). Hours with power shortages at the zone G target bus are shaded in red. In the top panel of (b), E-G interface utilization is uninformative from a prediction standpoint since the time series is almost constant. In the bottom panel of (b), E-G interface congestion is almost a perfect predictor of unmet demand at the target zone G bus, \emph{i.e.}, the time series contains no new predictive information beyond what is already contained in the past history of power shortages.}
    \label{fig:EGif_examples}
\end{figure}

\begin{figure}[hbtp]
    \centering
    \includegraphics[height=2.0in]{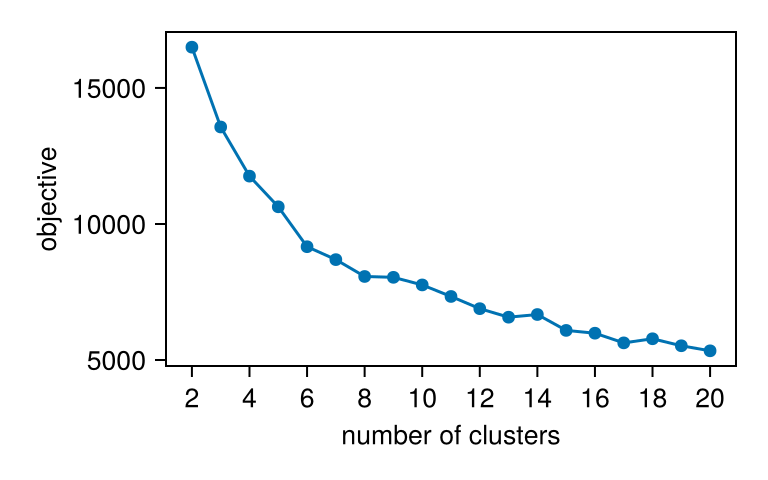}
    \caption{Within-cluster dissimilarity vs. the number of clusters for patterns of predictive transmission pressure points.}
    \label{fig:cluster_obj_vs_k}
\end{figure}

\begin{figure}[hbtp]
    \centering
    \makebox[\textwidth][c]{
        \includegraphics[height=3.0in]{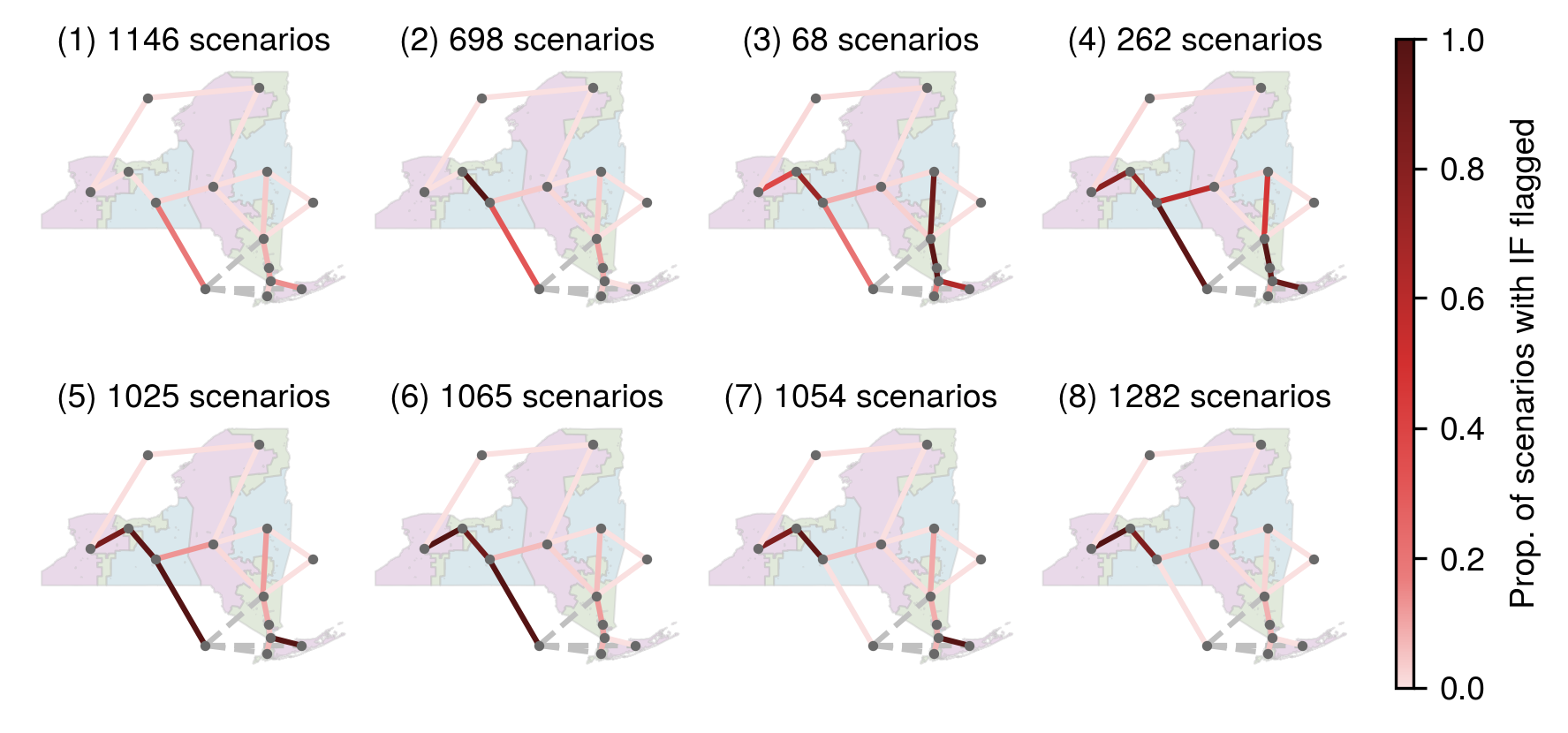}
    }
    \caption{Patterns of predictive transmission pressure points grouped into eight clusters. The size of each cluster is indicated in parentheses.}
    \label{fig:map_8_clusters}
\end{figure}

\begin{figure}[hbtp]
    \centering
    \includegraphics[width=\linewidth]{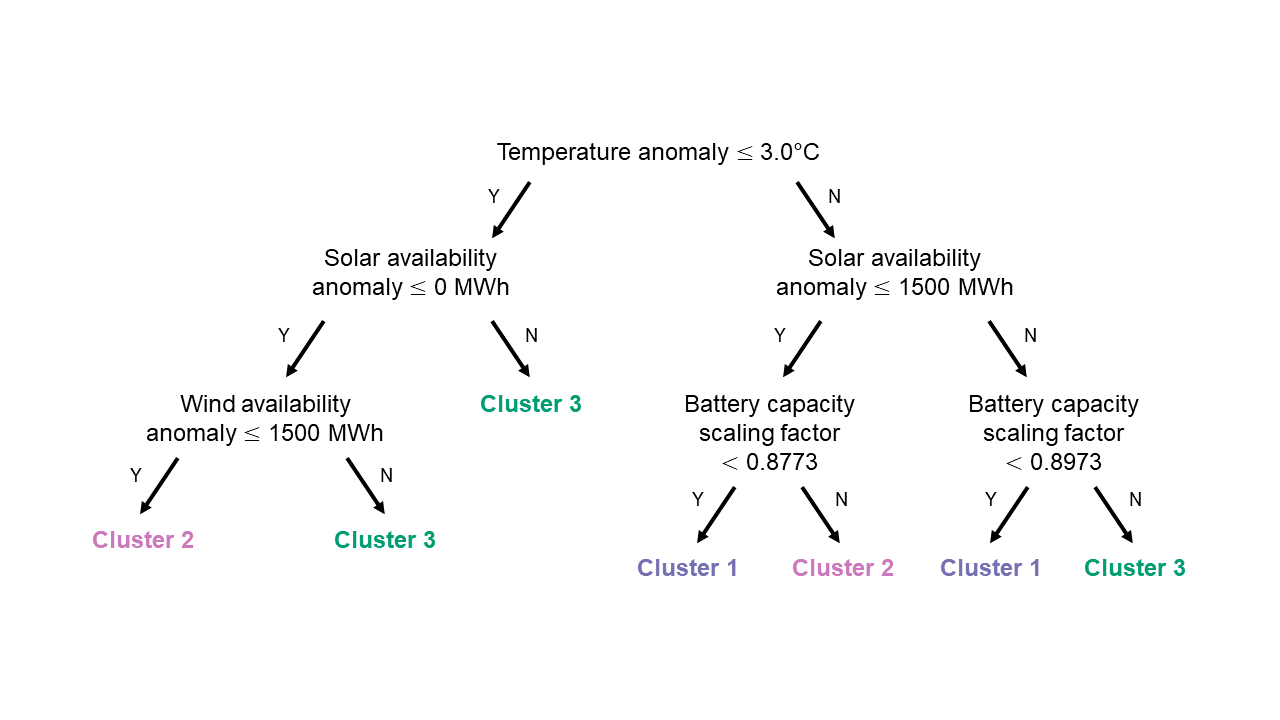}
    \caption{Classification tree to predict cluster membership based on meteorological-technological features of each scenario-year. The three clusters correspond to the clusters in Figure~\ref{fig:map_3_clusters} of the main text.}
    \label{fig:class_tree}
\end{figure}

\newpage
\subsection*{State space reconstruction for transfer entropy estimation}
Before estimating $TE(X\to Y),$ it is necessary to reconstruct the state space of the raw time series data $\{x_t\}_{t=1}^T$ and $\{y_t\}_{t=1}^T.$ 
The most common approach to state space reconstruction is to assume that the state of the source and target process at time $t$---say, $X_t$ and $Y_t,$ respectively---take the form of time-delay embedding vectors.\cite{wibral_transfer_2014}
That is, at time $t,$ we have
$$Y_t=\left(y_{t}, y_{t-\tau_Y}, \dots, y_{t-(d_Y-1)\tau_Y}\right)$$
and
$$X_t=\left(x_{t}, x_{t-\tau_X}, \dots, x_{t-(d_X-1)\tau_X}\right).$$
This leaves us with four relevant parameters to determine: the embedding dimensions, $d_X$ and $d_Y$, and the embedding delays, $\tau_X$ and $\tau_Y.$

This state space reconstruction step is critical to avoid under- or overestimating $TE(X\to Y)$.\cite{wibral_transfer_2014,bossomaier_introduction_2016} 
If past state variable $Y^-$ does not capture all the relevant history of process $Y,$ we risk overestimating $TE(X\to Y)$ by misattributing predictive information to process $X.$ 
That is, realizations of past state variables $Y_t^-$ should be maximally informative about the future $y_{t+1}$ of the target process.
Similarly, if $X^-$ does not capture all the relevant history of process $X,$ we risk underestimating $TE(X\to Y)$ by missing predictive information in process $X.$
Ragwitz and Kantz\cite{ragwitz_markov_2002} propose a method for identifying the optimal embedding dimension $d$ and delay $\tau$ by minimizing the prediction error of locally constant predictors in the time-delay embedding space; this yields time-delay embedding vectors that are optimally self-predictive.

\begin{figure}[hbtp]
    \centering
    \makebox[\textwidth][c]{
        \includegraphics[height=3.0in]{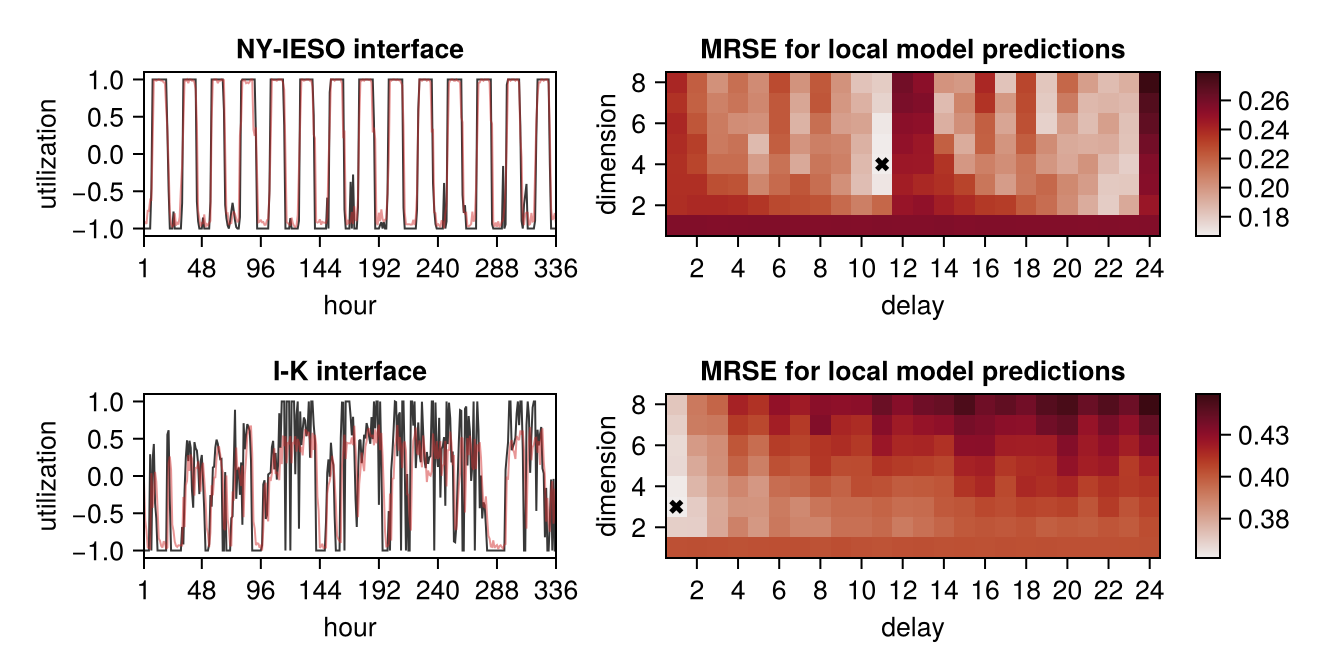}
    }
    \caption{Ragwitz criterion for identifying optimal embedding parameters. In the left column, the true time series of interface utilization is plotted in black, and the local model prediction---using optimal embedding parameters---is plotted in red. In the right column, MRSE for the local model is plotted for different combinations of embedding dimension and delay.}
    \label{fig:ragwitz}
\end{figure}

For time series with a strong periodicity, that period is reflected in the performance of different delay parameters; for time series with less structure, the optimal parameters are harder to predict. See Figure~\ref{fig:ragwitz}.

\subsection*{Surrogate time series generation}
\begin{figure}[hbtp]
    \centering
    \makebox[\textwidth][c]{
        \includegraphics[height=2.25in]{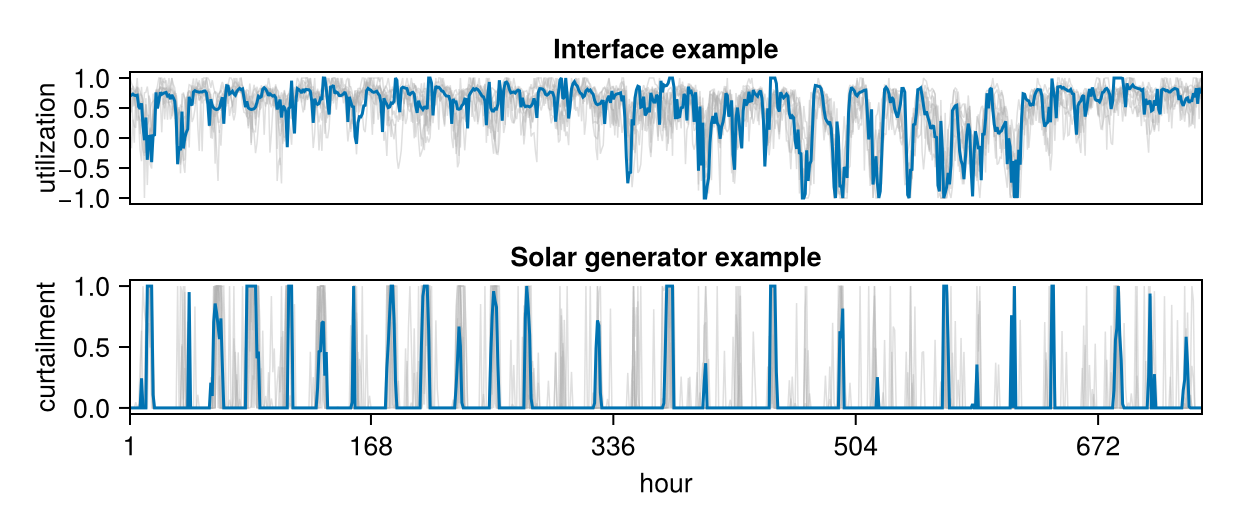}
    }
    \caption{Examples of surrogate time series generated using wavelet-based method from Keylock et al.\cite{keylock_constrained_2006}. The original series is plotted in blue, and ten surrogate time series are plotted in gray.}
    \label{fig:wls_surro_examples}
\end{figure}

To estimate the distribution of $TE(X\to Y)$ under the null hypothesis that there is no predictive information transfer from $X$ to $Y$, we wish to generate surrogate source time series $\{\hat{x}\}_{t=1}^T$ that preserve important statistical properties of the original source time series $\{x\}_{t=1}^T$ (e.g., autocorrelation properties) but break any temporal precedence structure between the source and target.
This can be naïvely achieved by randomly shuffling embedded source state space vectors $X^-$ in time or---in order to avoid making the surrogate time series more stationary than the original data---choosing a random cut point in the source time series and swapping the two halves.\cite{wibral_transfer_2014}
In our case, we wish to preserve the overall co-variability of our source processes with the occurrence of power shortages in zone G. 
This should prevent us from falsely flagging too many pressure points.
A wavelet-based surrogate generation method\cite{keylock_constrained_2006} allows us to constrain time domain behavior, thus preserving local mean and variance of the original time series (Figure~\ref{fig:wls_surro_examples}).

\end{document}